\documentclass[pdflatex,sn-mathphys-num]{sn-jnl}

\usepackage{graphicx}%
\usepackage{multirow}%
\usepackage{amsmath,amssymb,amsfonts}%
\usepackage{amsthm}%
\usepackage{mathrsfs}%
\usepackage[title]{appendix}%
\usepackage{xcolor}%
\usepackage{textcomp}%
\usepackage{manyfoot}%
\usepackage{booktabs}%
\usepackage{algorithm}%
\usepackage{algorithmicx}%
\usepackage{algpseudocode}%
\usepackage{listings}%

\usepackage[most]{tcolorbox} 
\usepackage{lipsum} 
\usepackage[utf8]{inputenc}
\usepackage[most]{tcolorbox}
\usepackage{xcolor}
\usepackage{multirow}
\usepackage{colortbl}
\usepackage{makecell}
\usepackage{enumitem} 
\usepackage{listings}
\usepackage{enumitem}
\usepackage{caption}
\usepackage{textcomp}

\definecolor{lavender}{rgb}{0.8, 0.7, 1}

\newcommand{\answer}[1]{
\noindent\fbox{%
    \parbox{.97\columnwidth}{%
        {#1}
    }%
}}

\newcommand{\nff}[1]{\textcolor{black}{#1}}

\newtcolorbox{modelbox}[1][]{%
    colback=gray!10,             
    colframe=black!100,            
    boxrule=0.1mm,               
    fonttitle=\bfseries,         
    coltitle=black,              
    enhanced,                    
    before skip=10pt,            
    after skip=10pt,             
    parskip=false,  
    attach boxed title to top left={xshift=3mm, yshift=-2mm, yshifttext=-1mm}, 
    boxed title style={%
        colback=green!60!black,  
        colframe=black!100, 
        boxrule=0.1mm,             
        arc=2pt,                 
        width=3cm,               
    },
    title={\textcolor{white}{\textbf{Input Prompt}}}, 
    #1
}

\newtcolorbox{modelboxoutput}[1][]{%
    colback=gray!10,             
    colframe=black!100,            
    boxrule=0.1mm,               
    fonttitle=\bfseries,         
    coltitle=black,              
    enhanced,                    
    before skip=3pt,            
    after skip=3pt,             
    attach boxed title to top left={xshift=3mm, yshift=-2mm, yshifttext=-1mm}, 
    boxed title style={%
        colback=orange!90!white,  
        colframe=black!100, 
        boxrule=0.1mm,             
        arc=2pt,                 
        width=3cm,               
    },
    title={\textcolor{white}{\textbf{Model Response}}}, 
    #1
}

\theoremstyle{thmstyleone}%
\newtheorem{theorem}{Theorem}
\newtheorem{proposition}[theorem]{Proposition}%

\theoremstyle{thmstyletwo}%
\newtheorem{example}{Example}%
\newtheorem{remark}{Remark}%

\theoremstyle{thmstylethree}%
\newtheorem{definition}{Definition}%

\begin{document}

\title[Article Title]{TVR: Automotive System Requirements \underline{T}raceability \underline{V}alidation and \underline{R}ecovery Through Retrieval-Augmented Generation}


\author[1]{\fnm{Feifei} \sur{Niu}}\email{feifeiniu96@gmail.com}

\author[1]{\fnm{Rongqi} \sur{Pan}}\email{rpan@uottawa.ca}

\author[1,2]{\fnm{Lionel C.} \sur{Briand}}\email{lbriand@uottawa.ca}

\author[3]{\fnm{Hanyang} \sur{Hu}}\email{phenom.hu@gmail.com}

\affil[1]{\orgname{University of Ottawa}, \city{Ottawa}, \country{Canada}}

\affil[2]{\orgname{Research Ireland Lero Centre, University of Limerick}, \city{Limerick}, \country{Ireland}}

\affil[3]{\orgname{Wind River Systems}, \country{Canada}}


\abstract{In automotive software development, as well as other domains, traceability between stakeholder requirements and system requirements is crucial to ensure consistency, correctness, and regulatory compliance. However, erroneous or missing traceability relationships often arise due to improper propagation of requirement changes or human errors in requirement mapping, leading to inconsistencies and increased maintenance costs. Existing approaches do not address traceability between stakeholder and system requirements and are not validated on industrial data, where engineers manually establish the links between requirements. 
Additionally, there are variations in how requirements are expressed, posing challenges for training-based approaches, particularly in large-scale and heterogeneous automotive systems.
Recent advancements in large language models (LLMs) provide new opportunities to address these challenges. In this paper, we introduce TVR, a requirement \underline{T}raceability \underline{V}alidation and \underline{R}ecovery approach primarily targeting automotive systems, leveraging LLMs enhanced with retrieval-augmented generation (RAG). TVR is designed to validate existing traceability links and recover missing ones with high accuracy. 
We empirically evaluate TVR on real automotive requirements, achieving 98.87\% accuracy in traceability validation and 85.50\% correctness in traceability recovery. Additionally, TVR demonstrates strong robustness, achieving 97.13\% in accuracy when handling unseen requirement variations. 
The experimental results highlight the practical effectiveness of TVR in industrial settings, offering a promising solution for improving requirements traceability in complex automotive systems.}




\maketitle

\section{Introduction}
Automotive systems have become increasingly complex, comprising a wide range of integrated components and features, including powertrain, chassis, infotainment, advanced driver assistance, and electric and autonomous driving systems. These components collaborate through electronic control units (ECUs), sensors, actuators, and software, working seamlessly to deliver performance, safety, and user experience~\cite{wang2024review}. 
To ensure high-quality software development, the automotive industry adheres to the Automotive Software Process Improvement and Capability dEtermination (ASPICE) standard~\cite{idri2016survey, garcia2012development}, which is based on the ISO 3300x norm group~\cite{ISO33001:2015, ISO33002:2015, ISO33003:2015, ISO33004:2015}. A fundamental aspect of ASPICE-compliant software development is requirements engineering, in which requirements are defined at multiple levels, including stakeholder and system requirements. System requirements are derived from stakeholder requirements, with traceability links established to ensure consistency and correctness, especially amid frequent changes.

Maintaining accurate and high-quality traceability links between stakeholder and system requirements is indeed crucial for multiple reasons: 1) \textbf{Improving system consistency}: Consistent traceability ensures that system requirements accurately reflect stakeholder needs, thus minimizing functional errors and inconsistencies~\cite{wiegers2013software, pargaonkar2023synergizing}. 2) \textbf{Preventing error propagation}: Detecting incorrect requirement mappings early reduces maintenance cost overheads and avoids large-scale rework~\cite{tufail2017systematic}. 3) \textbf{Ensuring regulatory compliance}, thus complying with ASPICE and functional-safety standards, including ISO~26262~\cite{siegl2010model}, to help improve the safety, reliability, and auditability of automotive software~\cite{qusef2011scotch}.

However, system engineers have reported challenges regarding traceability between requirements. 
One specific, particularly critical type of requirement we focus on here concerns Diagnostic Trouble Codes (DTCs), which are standardized vehicle error codes indicating malfunctions detected by the diagnostic system~\cite{marscholik2009road, pirasteh2019interactive}. While automotive diagnostic tools---with the market valued at USD 38.45 billion in 2023 and projected to reach USD 56.07 billion by 2031---play a crucial role in early issue detection, maintaining correct traceability of DTC requirements remains largely manual and labor-intensive. This limits the overall efficiency of the diagnostic process due to the time and effort required to manually verify the consistency between requirements and trace the system faults to requirements.

According to system engineers, these challenges, primarily due to resource and time constraints, include:
1) \textbf{Improper propagation of requirement changes}: When stakeholder requirements are modified (e.g., new requirements are added or existing ones are updated), the corresponding system requirements may not be updated accordingly, leading to outdated or invalid traceability links. If a requirement is deleted or merged, the original traceability link may persist, leading to incorrect mappings and potentially causing functional inconsistencies or safety risks, resulting in irrecoverable losses in safety-critical domains. 
For example, in 1998, NASA’s \$327 M Mars Climate Orbiter~\footnote{\url{https://en.wikipedia.org/wiki/Mars_Climate_Orbiter}} failed because mission directives mandated SI units, but the supporting software output imperial units. Without end-to-end traceability, the mismatch went undetected, causing trajectory errors and mission loss. This underscores the need for automated, robust requirement-consistency validation in industrial systems.
2) \textbf{Errors in requirement mapping}: Due to human error, system requirements may be incorrectly mapped to unrelated stakeholder requirements, or traceability links may be missing. According to our industry partner, they spend several weeks each year manually validating the traceability and consistency of requirements, as ensuring compliance with stakeholder requirements is always a top priority. Automating this process could reduce their manual effort and associated labor costs.

Unlike similarity-based approaches~\cite{gao2022using, hey2021improving, panichella2013and, kuang2015can, antoniol2002recovering, marcus2003recovering, keim2024recovering, keim2023detecting, cleland2005utilizing} that reconstruct links between requirements from scratch, industrial settings must manage preexisting traceability links, which, due to labeling errors or version changes, can incorrectly tie together unrelated or diverged requirements. Hence, it is essential to validate these links to ensure the connected requirements remain semantically and content-wise consistent.

The goal of this paper is to provide automated, effective ways to address the challenges above and thus identify \textbf{invalid and missing traceability links between DTC requirements}. Because requirements are not always expressed consistently within or across projects, we leverage large language models (LLMs), for their demonstrated robustness to linguistic variation~\cite{brown2020language, qin2023chatgpt}. 
Although prompt‐based LLM approaches have been recently applied to traceability recovery~\cite{fuchss2025lissa, rodriguez2023prompts}, they either rely solely on simple Zero-Shot or Chain-of-Thought (CoT) prompts (e.g., ``Is there a traceability link?'') for validating traceability links, which are inadequate for precisely validating the links between DTC system and stakeholder requirements that tend to slightly differ in the character strings of messages and signals.
 
To address the traceability challenges described above in the DTC requirements, we propose TVR, a Retrieval-Augmented Generation (RAG) approach that leverages generative LLMs to support traceability validation and recovery. Unlike existing RAG-based traceability recovery approaches (e.g., \cite{fuchss2025lissa, rodriguez2023prompts}), which primarily rely on retrieval to compute similarity scores between software artifacts to determine whether a traceability link exists, TVR adopts a fundamentally different approach to RAG. Specifically, TVR retrieves similar requirement pairs, including both positive (valid) and negative (invalid) traceability examples, and incorporates them into the prompt to explicitly guide and ``teach'' the LLM how to reason about traceability correctness.
\nff{While traditional RAG systems are primarily designed for knowledge-intensive tasks, where external information must be retrieved from documents, knowledge bases, or repositories to provide necessary domain knowledge in the language model~\cite{lewis2020retrieval}, requirements traceability validation differs fundamentally from such tasks. The challenge is not the absence of domain knowledge, but rather determining whether a stakeholder requirement and a system requirement exhibit a valid traceability relationship despite differences in terminology, abstraction level, and specification style. In our industrial context, traceability validation primarily relies on identifying recurring correspondence patterns between stakeholder and system requirements, rather than on retrieving additional information from external documentation. Therefore, instead of retrieving external knowledge, TVR retrieves previously labeled requirement pairs and uses them as in-context demonstrations. These examples provide concrete decision patterns that help the LLM understand how traceability correctness is assessed in practice. Consequently, TVR can be viewed as a retrieval-augmented in-context learning framework, in which retrieval augments reasoning with representative examples rather than augmenting knowledge with external documents.
}

We investigate and leverage 13 LLMs to validate the correctness of traceability links and further recover missing links between requirements. We evaluated TVR on 2,132 DTC requirement pairs from an automotive system, achieving an overall accuracy of 98.87\%. When handling different requirements variants, TVR still maintains an accuracy of 97.13\%, demonstrating strong robustness. Additionally, TVR successfully identified 502 missing traceability links with 85.50\% accuracy within the dataset.

Our contributions include:
\begin{itemize}

   \item We propose TVR, a RAG-enhanced LLM approach for validating traceability links between automotive stakeholder and system requirements.
   \item We evaluate and demonstrate the robustness of TVR on unseen requirement variations, as requirements writing conventions often vary in industrial contexts.
   \item We also apply TVR to recover traceability links, achieving high accuracy.
   \item We comprehensively investigate 13 LLMs across four prompting strategies and our RAG-based TVR approach, and examine the impact of different similarity measures and the number of examples on TVR performance. The implementation is made publicly available~\cite{TVR2025}.
\end{itemize}

The remainder of this paper is structured as follows: Section~\ref{sec:background} defines the industrial problems addressed and provides the necessary context. Section~\ref{sec:approach} introduces our TVR approach. The study design is detailed in Section~\ref{sec:studydesign}, followed by an analysis of the experimental results in Section~\ref{sec:results}, and a discussion in Section~\ref{sec:discussion}. Section~\ref{sec:threats} examines threats to validity. Section~\ref{sec:relatedwork} reviews the state of the art in requirements traceability research and contrasts it with our contributions. Finally, we conclude the paper in Section~\ref{sec:conclusion}.

\section{Problem Definition and Challenges}\label{sec:background}
This work aims to support traceability validation and recovery between stakeholder requirements and system requirements for DTCs in automotive systems. This section provides an overview of DTCs and requirements traceability, with a particular focus on stakeholder and system requirements. Although our terminology and sanitized examples originate in the automotive domain, many other critical domains that typically require compliance with functional safety standards face similar challenges and concepts. 

\subsection{Diagnostic Trouble Code Requirements}\label{sec:dtc}
\subsubsection{Diagnostic Trouble Code}
Diagnostic trouble codes (DTCs), also known as fault codes, are standardized codes used in automotive systems to identify and diagnose issues within a vehicle's Electronic Control Units (ECUs)~\cite{marscholik2009road, pirasteh2019interactive}. These codes are generated when the On-Board Diagnostics (OBD) system detects a malfunction in components such as the engine, transmission, or emission systems. Example conditions that trigger a DTC are implausible or erroneous signal values or signals that were not received~\cite{theissler2017multi}.
DTCs typically consist of a letter (indicating the system, e.g., P for Powertrain) followed by four digits that provide specific details about the fault. Technicians and diagnostic tools use these codes to pinpoint problems efficiently, aiding in vehicle repair and maintenance.

While DTC codes themselves are standardized and follow well-defined diagnostic categories, the associated DTC behavior---including detection conditions, signal dependencies, plausibility checks, timing thresholds, and system responses---is not standardized and is instead defined in the corresponding requirements. In modern automotive systems, DTC requirements are expressed at various levels of abstraction: stakeholder, system, software, and so on. These requirements define the conditions under which faults are detected, recorded, and communicated within ECUs, ensuring that the system operates effectively and meets both technical and regulatory standards. In this study, we focus on the requirements pertaining to the \textit{setting} and \textit{clearing} of DTCs, corresponding to the conditions for so-called \textit{mature} and \textit{demature} DTCs, respectively, which are further described in Section~\ref{sec:dtc_system_req}.

\subsubsection{DTC Stakeholder Requirements}
Stakeholder requirements are the high-level needs and expectations of all parties (drivers, mechanics, regulators, manufacturers, etc) involved or affected by the automotive system. As shown in Figure~\ref{fig:stakereq}, DTC stakeholder requirements include, but are not limited to, the following key elements:

\noindent
\textbf{Trigger Condition:} Specify the conditions under which a DTC is set or cleared.  

\noindent
\textbf{Input Message:} Define the message that triggers the setting or clearing of the DTC.  

\noindent
\textbf{Mitigation Action:} Include specific actions for setting the DTC to ``Present'' or ``Not Present''.  

\noindent 
\textbf{Validation Rules:} Establish validation rules to ensure that DTC setting and clearing operations comply with design standards. 

\noindent
\textbf{Reference Documents:} Provide relevant standards and documents supporting the setting and clearing of the DTC. 

For example, VARIATION 1 in Figure~\ref{fig:stakereq} describes a stakeholder requirement specifying how a standardized ``Lost Communication'' DTC should be handled in a concrete system context. In this requirement, $MESSAGE\_1$ refers to a specific cyclic (or cyclic-on-change) communication message that module $M$ is expected to periodically receive from a source ECU during normal operation. A Lost Communication DTC is triggered when such an expected message is absent for multiple consecutive communication cycles, corresponding to a failure mode commonly referred to as a missing message. Accordingly, if $MESSAGE\_1$ is not received within a predefined number of consecutive message cycles, the stakeholder requirement specifies the intended diagnostic outcome: setting the DTC state according to predefined rules.

However, the examples in Figure~\ref{fig:stakereq} illustrate only four representative DTC stakeholder requirements, selected to exemplify common patterns of variation observed across the industrial dataset. In practice, DTC stakeholder requirements vary widely in terminology, abstraction levels, and specification structures. These variations undermine the assumptions of consistency and comparability required by traditional traceability approaches. As a result, keyword- and rule-based methods become brittle, and learning-based approaches struggle to generalize across heterogeneous stakeholder specifications, making reliable traceability between stakeholder requirements and system-level specifications challenging.

\begin{figure*}[!htbp]
    \centering
    \begin{tcolorbox}[
      colframe=lavender!100!black,
      colback=lavender!10,
      title={Stakeholder Requirements},
      fonttitle=\bfseries\small,   
      fontupper=\small             
    ]
    \noindent   
    \fcolorbox{lavender!100!black}{lavender!100}{\textbf{VARIATION 1}}
         If \textit{Trigger\_Condition} = ``RUN'', and the module \textit{M} does NOT receive the message \textit{MESSAGE\_1} for a certain number of message cycles, then the module \textit{M} shall:
   --- Set the DTC to ``Present'' according to rules contained in the ``\textit{Reference\_Document}''. \fcolorbox{gray!70!black}{gray!30}{\textbf{Lost Communication, Mature}}

    \noindent
    \fcolorbox{lavender!100!black}{lavender!100}{\textbf{VARIATION 2}} 
    If \textit{Trigger\_Condition} = ``RUN'', the internal signal \textit{MODULE\_MODE} != ``\textit{SIGNAL\_1\_Fail}'', then the module \textit{M} must:  
 ---    Set the DTC to ``Not Present'' according to the appropriate validation rules contained in the ``\textit{Reference\_Document}''. \fcolorbox{gray!70!black}{gray!30}{\textbf{Implausible Data, Demature}}
     
    \noindent
    \fcolorbox{lavender!100!black}{lavender!100}{\textbf{VARIATION 3}}
    If \textit{Trigger\_Condition} = ``RUN'', and the module \textit{M} does not detect a Plausibility Fault on a signal within the message ``\textit{MESSAGE\_2}'', then the module \textit{M} must:  
 --- Set the DTC to ``Not Present'' according to rules contained in the ``\textit{Reference\_Document}''.  \fcolorbox{gray!70!black}{gray!30}{\textbf{Implausible Data, Demature}}

    \noindent
    \fcolorbox{lavender!100!black}{lavender!100}{\textbf{VARIATION 4}}
    If \textit{Trigger\_Condition} = ``RUN'', the module \textit{M} determines there is a failure in \textit{SIGNAL\_2}, then the module \textit{M} shall:
    --- Set the \textit{INTERNAL\_FLAG} = ``Faulted''. 
    Set the DTC to ``Present'' according to rules contained in the ``\textit{Reference\_Document}''.   \fcolorbox{gray!70!black}{gray!30}{\textbf{Implausible Data, Mature}}
    \end{tcolorbox}
    \caption{Fictitious Variations of Stakeholder Requirements (Due to data privacy concerns, only fictitious examples are shown here.
Italicized variables in the examples (e.g., \textit{MESSAGE\_1}) are pseudonyms---real values were employed in the experiments).}
    \label{fig:stakereq}
\end{figure*}

\subsubsection{DTC System Requirements}
\label{sec:dtc_system_req}

System requirements define the detailed specifications that the automotive system must meet to satisfy stakeholder requirements. These requirements translate stakeholder expectations into specific, actionable objectives for the system.  

In our context, the elements of a DTC system requirement include \textit{Name}, \textit{Number}, \textit{Description}, \textit{Priority}, \textit{Enable Condition}, \textit{Mature}, \textit{Mature Time}, \textit{Demature}, \textit{Demature Time}, and others. Among these, the most critical elements are \textit{Mature} and \textit{Demature}. \textit{Mature} specifies the conditions that must be met for a DTC to be set, while \textit{Demature} defines the conditions required for the DTC to be cleared. \textit{Mature} and \textit{Demature} should be negations of each other.  

Figure~\ref{fig:example} presents an example of a sanitized \textit{Mature} condition from a system requirement, while \textit{Demature} follows a similar structure.
Both \textit{Mature} and \textit{Demature} processes start by verifying that the relevant components are enabled. This ensures that the system is actively monitoring the specified components and is ready to conduct further checks if they are available and operational. Next, the system verifies proper communication and ensures that the expected data has been received, confirming that the system is functioning correctly. If a malfunction is detected, the system sets the corresponding DTCs; otherwise, the DTCs are cleared.


\begin{figure}
  \centering
  \begin{tcolorbox}
  [colframe=green!50!black, colback=green!10, title=System Requirement
  ]
    \begin{lstlisting}[basicstyle=\ttfamily\scriptsize]
if ( ENABLE_COMPONENT is enabled ){ 
    if ( Missing_Msg_MESSAGE_1 || 
         Missing_Msg_MESSAGE_2 || ... ){ 
        LostComm_Module_M = TRUE; }}
    \end{lstlisting}
  \end{tcolorbox}
  \caption{Example of Sanitized DTC System Requirement \nff{Illustrating a Mature Condition for Setting a DTC}. Note: Real values were employed in the experiments.}
  \label{fig:example}
\end{figure}

\subsubsection{DTC Types} 
There are different types of ECU failures, corresponding to various DTC types, which can be broadly categorized into \textit{internal} and \textit{network} failures~\cite{palai2013vehicle, theissler2017multi}.
In this study, we focus on two common types of network failures~\cite{palai2013vehicle, theissler2017multi}:
\begin{itemize}
    \item Lost Communication: This occurs when an ECU fails to receive an expected message from a source ECU. The absence of an expected message is categorized as a \textit{Missing Message} failure mode.
    
    \item Implausible Data: This occurs when an ECU receives an expected message but detects untrustworthy data that is inconsistent, unrealistic, or beyond the expected domain. 
\end{itemize}


\nff{Although different DTC categories correspond to different fault scenarios, their associated system requirements share a common structure, typically including elements such as~\textit{Name}, \textit{Number}, \textit{Description}, \textit{Priority}, \textit{Enable Condition}, \textit{Mature}, \textit{Mature Time}, \textit{Demature}, and \textit{Demature Time}, and others. Among these elements, \textit{Mature} and \textit{Demature} are the most relevant for traceability. A stakeholder requirement is traced to either a \textit{Mature} condition or a \textit{Demature} condition of the corresponding system requirement. Specifically, a stakeholder requirement whose mitigation action is ``Set the DTC to Present'' corresponds to a \textit{Mature} condition, whereas a stakeholder requirement whose mitigation action is ``Set the DTC to Not Present'' corresponds to a \textit{Demature} condition.}

\begin{figure}
    \centering
    \includegraphics[width=0.9\linewidth]{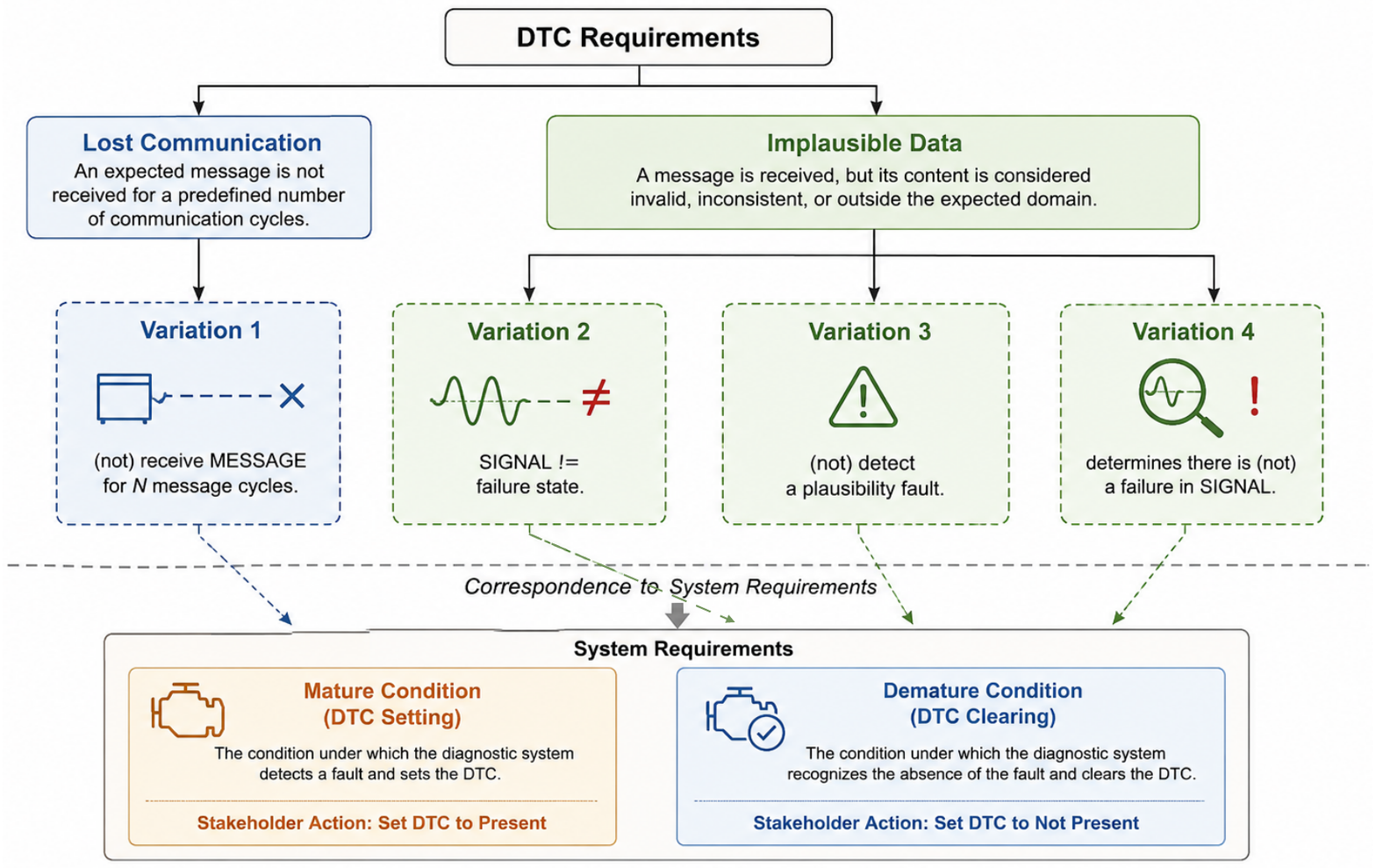}
    \caption{Taxonomy of stakeholder requirement variations for DTC requirements and their correspondence to system requirements.}
    \label{fig:category}
\end{figure}

\nff{To systematically characterize stakeholder requirements, we categorize them according to how the diagnostic condition is formulated in the requirement text. Figure~\ref{fig:category} presents the resulting taxonomy. Based on the stakeholder requirement formulations observed in our industrial dataset, we identified four recurring variations, which we describe in detail below.}

\nff{\textbf{VARIATION 1 (Lost Communication).} This variation corresponds to requirements describing ``Lost Communication''. The diagnostic condition is expressed in terms of whether a communication message is received, typically formulated as ``the module receives (or does not receive) \textit{MESSAGE} for a certain number of message cycles''. In this variation, receiving the message corresponds to a \textit{Demature} condition, whereas not receiving the message corresponds to a \textit{Mature} condition.}

\nff{\textbf{VARIATION 2 (Implausible Data).} This variation corresponds to requirements describing ``Implausible Data''. The diagnostic condition is expressed as a comparison between an internal signal and a predefined failure state. Typical formulations include expressions such as ``\textit{MODULE\_MODE} = \textit{SIGNAL\_1\_Fail}'' or ``\textit{MODULE\_MODE} != \textit{SIGNAL\_1\_Fail}''. In this variation, equality with the failure state corresponds to a \textit{Mature} condition, whereas inequality corresponds to a \textit{Demature} condition.}

\nff{\textbf{VARIATION 3 (Implausible Data).} This variation corresponds to requirements describing ``Implausible Data''. The diagnostic condition is expressed in terms of whether a plausibility fault is detected. Typical formulations include expressions such as ``the module detects (or does not detect) a plausibility fault on a signal within \textit{MESSAGE\_2}''. In this variation, detecting a plausibility fault corresponds to a \textit{Mature} condition, whereas not detecting a plausibility fault corresponds to a \textit{Demature} condition.}

\nff{\textbf{VARIATION 4 (Implausible Data)}. This variation corresponds to requirements describing ``Implausible Data''. The diagnostic condition is expressed through an explicit determination of signal failure. Typical formulations include expressions such as ``the module determines there is (or is not) a failure in \textit{SIGNAL\_2}''. In this variation, determining the existence of a failure corresponds to a \textit{Mature} condition, whereas determining the absence of a failure corresponds to a \textit{Demature} condition.}

\nff{These four variations cover all stakeholder requirement formulations observed in the industrial dataset used in this study. The variations were identified through an iterative manual analysis of all stakeholder requirements by two annotators, who grouped them according to how their diagnostic conditions are formulated. Although additional variations may appear in other projects or larger datasets, all stakeholder requirements in our studied dataset can be assigned to one of these four variation categories.}

\nff{Stakeholder requirements may differ in their module names, message names, signal names, parameter values, or other domain-specific details. However, requirements sharing the same diagnostic condition formulation are assigned to the same variation category. Consequently, each stakeholder requirement belongs to exactly one variation, while multiple stakeholder requirements may instantiate the same variation using different domain-specific entities.}

\nff{Figure~\ref{fig:stakereq} illustrates sanitized examples of the four stakeholder requirement variations identified in our dataset. Variation 1 corresponds to ``Lost Communication'' requirements, whereas Variations 2–4 correspond to different formulations of ``Implausible Data'' requirements. Depending on the specific diagnostic condition being expressed, a stakeholder requirement belonging to a given variation may correspond to either a \textit{Mature} condition (setting the DTC to ``Present'') or a \textit{Demature} condition (setting the DTC to ``Not Present'') in the linked system requirement.}

\subsection{Requirements Traceability Challenges}

Requirements traceability refers to the ability to track and document the lifecycle of a requirement---both forward and backward---from its origin through development, implementation, and usage~\cite{gotel1994analysis}. In the context of automotive systems, traceability helps ensure that stakeholder needs are effectively aligned with the system design. It plays a critical role in confirming that the system meets customer expectations and regulatory standards. Moreover, traceability provides a clear audit trail for quality assurance, enabling the detection of gaps, inconsistencies, or changes in requirements throughout the development process.  

As discussed in Section~\ref{sec:dtc}, DTC requirements come in different types, each following distinct patterns. Moreover, both stakeholder and system requirements vary in how they are expressed, and unforeseen changes are possible in the future. This presents a significant challenge: \textbf{A general solution capable of handling diverse and unseen variations is essential}. Traditional NLP approaches~\cite{rahimi2018evolving, tufail2017systematic, charalampidou2021empirical}, based on supervised learning of observed variations, are challenged by unseen variations. Moreover, a comparison of stakeholder requirements (Figure~\ref{fig:stakereq}) and system requirements (Figure~\ref{fig:example}) raises another challenge: \textbf{Our traceability validation problem cannot be achieved by simply calculating text similarity or finding word overlaps}. Since the system requirement involves checking whether multiple messages or signals are missing, whereas a stakeholder requirement addresses only the handling of a single message, a system requirement may be traced to several stakeholder requirements, each covering a message or signal in the system requirement. Furthermore, as shown in Figure~\ref{fig:stakereq} and Figure~\ref{fig:example}, the differences between messages and signals at the character level are not significant; however, \textbf{a large number of domain-specific terminologies exist} in the description of different signals and messages.

To address these challenges, this paper proposes TVR, an approach for validating the traceability of requirements in automotive systems. TVR leverages the exceptional natural language understanding capabilities of LLMs. Instead of relying on the general concept of traceability links~\cite{guo2024natural}, we ask LLMs to validate whether a system requirement covers the message or signal specified in a stakeholder requirement. We guide the LLM to focus specifically on the message or signal rather than on other parts of the requirement to avoid confusion.

\section{TVR}\label{sec:approach}
\nff{TVR is a retrieval-augmented approach that leverages generative LLMs for requirements traceability validation and recovery. Unlike conventional RAG systems, which retrieve external documents to provide additional domain knowledge, TVR retrieves previously labeled requirement pairs as task-specific reasoning examples. This design is motivated by the nature of the traceability validation problem in our context: determining whether a stakeholder requirement is correctly covered by a system requirement does not require additional external knowledge; rather, it requires understanding recurring correspondence patterns between requirements. Therefore, instead of retrieving documents from an external knowledge base, TVR retrieves similar requirement pairs whose traceability status has already been validated by engineers and uses them as in-context demonstrations. These demonstrations provide concrete decision patterns that guide the LLM in assessing the correctness of traceability and help it generalize to previously unseen requirement pairs.}

Formally, given a language model \( G \), a training set  
\[
D_{\text{train}} = \left\{
\begin{aligned}
    &\langle stakeReq_i, sysReq_i, label_i \rangle \mid 
    i = 1, 2, \dots, N, \\
    &\text{where } label_i \in \{Yes, No \}
\end{aligned}
\right\}
\]  
and a test case  
\[
x_{\text{test}} = \langle stakeReq_{\text{test}}, sysReq_{\text{test}} \rangle,
\]  
TVR retrieves similar examples from \( D_{\text{train}} \) to assist \( G \) in predicting the label for \( x_{\text{test}} \).  

\begin{figure}[htbp]
\centering
\includegraphics[width=0.9\linewidth]{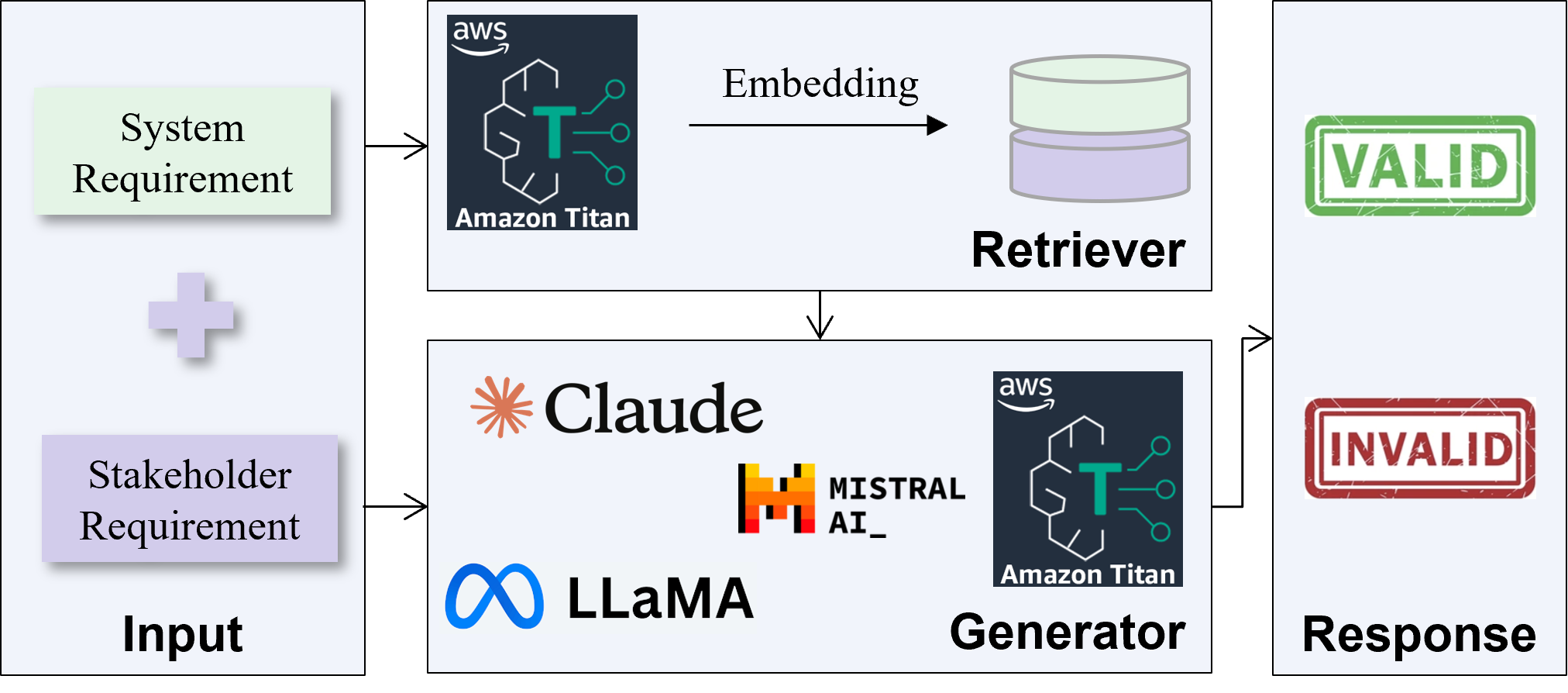}
\caption{Overall Framework of TVR.}
\label{fig:approach}
\end{figure}

The overall workflow of TVR, illustrated in Figure~\ref{fig:approach}, consists of two key components: a \textbf{retriever} and a \textbf{generator}. The retriever identifies the most similar examples from \( D_{\text{train}} \) based on their similarity to \( x_{\text{test}} \) and provides them as input to the generator. The generator then utilizes these retrieved examples to assess the validity of the traceability link between $stakeReq_{\text{test}}$ and $sysReq_{\text{test}}$.

\subsection{Retriever}

The retriever identifies and retrieves the most similar examples from \( D_{\text{train}} \) that closely match \( x_{\text{test}} \), serving as input to the generator.
The retriever operates in two stages. First, it generates embeddings for each data point (\( \langle \emph{stakeReq}, \emph{sysReq} \rangle \)) in \( D_{\text{train}} \) as well as for \( x_{\text{test}} \). To this end, the stakeholder requirement text and the corresponding system requirement text are concatenated using a whitespace separator, and a single embedding is generated for the resulting text. Then, using a similarity-based retrieval mechanism, it selects the \( k \times 2 \) most similar data points from \( D_{\text{train}} \), comprising \( k \) \emph{valid} and \( k \) \emph{invalid} examples.

Due to data privacy constraints, we utilize Amazon Titan~\cite{aws_titan}, the only model provided by our industry partner, to obtain embeddings for the concatenated requirement pairs \( \langle \emph{stakeReq}, \emph{sysReq} \rangle \). \nff{Although the choice of embedding model may influence which examples are retrieved, we do not expect TVR to be highly sensitive to this configuration. The retriever is used only to select representative in-context demonstrations, while the final traceability decision is still made by the LLM based on the retrieved examples and the input requirement pair. Therefore, moderate variations in the retrieved examples are unlikely to fundamentally alter the approach's behavior, as long as the embedding model can capture basic semantic similarity between requirement pairs.} We then leverage the FAISS library~\cite{faiss} to compute cosine similarity and efficiently retrieve the Top-\textit{k} most similar data points from both valid and invalid examples.

\nff{
All embeddings are generated offline and indexed in FAISS prior to retrieval. In our experimental setting, the retrieval repository consists of manually validated traceability links and remains fixed throughout each leave-one-out evaluation iteration. In industrial deployment scenarios, newly validated traceability links can be embedded using the same embedding model and incrementally added to the FAISS index, without requiring any modification to the generator component. Since new and existing examples are represented in the same embedding space, the retrieval repository can evolve together with the requirements database while allowing newly added examples to be searched alongside previously indexed ones.}

\nff{Regarding scalability, the retrieval complexity is dominated by nearest-neighbor search in the embedding space. FAISS is specifically designed for efficient similarity search over large vector collections and supports scalable indexing strategies for substantially larger repositories. Although our industrial dataset contains 2,132 traceability links, the retrieval mechanism is independent of the generator and can accommodate larger repositories as additional validated traceability links become available.
}

\subsection{Generator}
The generator plays a crucial role in evaluating the traceability between a given pair \( \langle stakeReq_\text{test}, sysReq_\text{test} \rangle \). This process leverages retrieval-augmented, in-context learning powered by LLMs, e.g., Claude 3.5 Sonnet~\cite{aws_anthropic}, the best model as presented in Section~\ref{sec:results}. It uses the top \(k \times 2\) most similar examples obtained by the retriever. These retrieved examples are incorporated into the generator’s prompt as contextual examples to guide its response generation. Using this augmented prompt, the generator analyzes the relationship between the stakeholder requirement ($stakeReq_\text{test}$) and the system requirement ($sysReq_\text{test}$). It generates a response indicating whether their traceability link is valid.

As shown in Figure~\ref{fig:prompt}, the prompt of the generator includes: 

\noindent
\textbf{1) Instruction:} The specific task that the model needs to perform.

\noindent
\textbf{Rationale:} We define our task as validating whether a system requirement \textit{covers} a stakeholder requirement. While \textit{traceability validation} is a broad concept, we narrow its scope in our context to focus specifically on whether the system requirement covers the message or signal in the stakeholder requirement. This refinement is necessary because a single system requirement may correspond to multiple stakeholder requirements in our case.
    
\noindent
\textbf{Prompt:} See Figure~\ref{fig:prompt}, Lines 1–2.
    
\noindent
\textbf{2) Context}: External information or additional context that can steer the model to better responses. 

\noindent
\textbf{Rationale:} In our context, we provide the \( k \times 2 \) most similar examples, including \( k \) valid and \( k \) invalid ones, to help LLMs learn to understand them.
    
\noindent
\textbf{Prompt:} See Figure~\ref{fig:prompt}, Line 3.
        
\noindent
\textbf{3) Input Data:} the input or question we aim to answer.  

\noindent
\textbf{Rationale:} We use XML tags to clearly separate different parts of the input data and ensure the prompt is well structured~\cite{claude_prompt}.

\noindent
\textbf{Prompt:} See Figure~\ref{fig:prompt}, Lines 4–7.
        
\noindent
\textbf{4) Output Indicator:} The type or format of the output.

\noindent
\textbf{Rationale:} For experimental purposes, we only require the model to output a predicted label, so the model only needs to respond with ``Yes'' or ``No.'' Note that in practice, if an explanation from LLMs is required, this sentence should be adjusted to enable LLMs to generate a step-by-step reasoning and analysis process.

\noindent
\textbf{Prompt:} ``only respond with either `Yes' or `No'.'' (Line 4)

With the above prompt, the generator analyzes the input requirement pair according to the instructions and provided contextual examples, and then responds with either ``Yes'' or ``No'' for the given pair \( \langle stakeReq_\text{test}, sysReq_\text{test} \rangle \).

\begin{figure}
\begin{modelbox} 
\textit{Line 1: }Please check if the message or signal from the stakeholder requirement is correctly \textbf{covered} by the system requirement. \\
\textit{Line 2: }Please focus only on verifying the \textbf{message} or \textbf{signal} mentioned, without considering other parts of the requirement.\\
\textit{Line 3: }Example: $<$example$>$\{\textit{k $\times$ 2 examples}\}$<$/example$>$\\
\textit{Line 4: }Now evaluate the following step by step and only respond with either ``Yes'' or ``No'':\\
\textit{Line 5: }Stakeholder Requirement: $<$stakeholder$>$ \{\textit{stakeReq}\} $<$/stakeholder$>$\\
\textit{Line 6: }System requirement: $<$system$>$ \{\textit{sysReq}\}$<$/system$>$\\
\textit{Line 7: }Response:
\end{modelbox}
\captionof{figure}{Input Prompt.}\label{fig:prompt}
\end{figure}

\subsection{Traceability Recovery} \label{sec:retrieving}
To recover potentially missing traceability links between stakeholder and system requirements in the dataset, we first pair every \emph{stakeReq} with every \emph{sysReq} that lacks a trace link. This results in several pairs approaching the Cartesian product of the two sets (excluding existing traceability links), which would be computationally expensive and time-consuming to analyze.  

To address this challenge, we apply the following preprocessing steps to reduce computational overhead:

\noindent
\textbf{Step 1. Variation Matching:} As discussed in Section~\ref{sec:background}, in the studied dataset, stakeholder requirements can be grouped into four distinct variations, each following a specific template, while system requirements are classified into two categories characterized by unique syntactic patterns. A traceability link can only exist between requirements that belong to the same DTC type. For example, a stakeholder requirement for ``Lost Communication'' can only be linked to a system requirement of the same DTC type. By excluding cross-category mismatches, we substantially reduce the number of pairs being considered.

\noindent
\textbf{Step 2. Condition Matching:} Each atomic stakeholder requirement corresponds exclusively to either a mature condition or a demature condition in the system requirement. Based on this distinction, we group \emph{stakeReq} and \emph{sysReq} accordingly and only match those within the same condition type, thus further eliminating irrelevant pairs.

\noindent
\textbf{Step 3: Message Overlap Matching:} If a stakeholder requirement and a system requirement do not share any message overlap, there is no traceability between them. To check this, we first tokenize the stakeholder and system requirements, remove stop words (customized for our domain), and extract the messages. We then compare these messages and retain only the pairs that share at least one common message. This step effectively reduces the number of pairs requiring further validation.
   

Through these three steps, we effectively reduce the number of candidate pairs while preserving all potential missing links. The filtered pairs are then used to validate the following hypothesis (\textbf{H}):  

\begin{quote}  
A valid traceability link, denoted as \emph{traceLink}, exists for \( \langle \emph{stakeReq}, \emph{sysReq} \rangle \).  
\end{quote}  

We use TVR to validate this hypothesis. If \textbf{H} is confirmed for a pair, it implies that the \emph{traceLink} is missing and should be added to the dataset. Conversely, if \textbf{H} is not confirmed, it indicates that no traceability link exists between \emph{stakeReq} and \emph{sysReq}. Through this process, we systematically recover missing traceability links in the dataset.  

\section{Study Design}\label{sec:studydesign}
In this section, we present our research questions, the LLMs and prompt engineering strategies evaluated, the dataset used to validate and recover requirements traceability, and the evaluation metrics used to assess TVR's performance.

\subsection{Research Questions}
\begin{enumerate}[label=RQ\arabic*:]
    \item What is the performance of LLMs with Zero-Shot prompting on requirements traceability validation? 

    This RQ aims to evaluate and compare the performance of available LLMs, including Llama, Claude, Titan, and Mistral, with Zero-Shot prompting.
    
    \item What is the best prompt strategy using the best LLM for requirements traceability validation? 
    
    This RQ investigates and compares the performance of various prompt engineering strategies (i.e., Zero-Shot, CoT, Few-Shot, and self-consistency) and the RAG-based TVR approach for requirements traceability validation.
    
    \item How robust is TVR for requirements traceability validation in the presence of unseen requirement variations?
    
  One of our main motivations in relying on LLMs is that we expect them to be more robust to unseen variations. Given the numerous variations that typically occur in stakeholder requirements, this RQ primarily evaluates TVR's performance under requirements. 
  In practice, this is important as we expect to continuously encounter new variations.
    
    \item What is the performance of TVR on traceability link recovery between stakeholder requirements and system requirements?

   LLMs can be used not only for traceability validation but also for missing-link recovery by determining whether valid traceability exists between any two requirements that are not linked. This research question leverages our TVR approach to recover missing links and evaluates its accuracy.   
\end{enumerate}

\subsection{Models and Prompts}

\subsubsection{LLMs}
We conducted experiments with 13 LLMs that were accessible to our industry partner at the time of the study through the Amazon Bedrock service, including Claude (Claude 3.5 Sonnet, Claude 3 Sonnet, Claude 2, Claude Instant, and Claude 3 Haiku), Llama (Llama 3 8B and Llama 3 70B), Mistral (Mistral 7B, Mixtral 8x7B, and Mistral Large 2402), and Titan (Titan Text Premier, Titan Text Express, and Titan Text Lite) (as listed in Table~\ref{tab:results}).
While we acknowledge the release of newer models in recent months, the 13 models evaluated in this study were the models made available to us by our industry partner at the time of experimentation. More recent models would probably yield even better results, but our conclusions would not be affected.

\subsubsection{Prompt Engineering}

We evaluated LLMs using several SOTA prompting strategies (i.e., Zero-Shot, Few-Shot, and CoT) and our RAG-based TVR approach. These strategies represent progressively advanced prompting techniques, ranging from simple direct inference (Zero-Shot) to more sophisticated approaches that incorporate reasoning steps, in-context examples, majority voting, and retrieval of relevant external information. The detailed prompts used in our experiments are provided in the replication package~\cite{TVR2025}.

\subsection{Dataset}

To evaluate the performance of LLMs for automotive requirements traceability validation and recovery, we used a dataset of DTC requirements from our industry partner. \nff{The dataset was collected from a single industrial automotive project and contains stakeholder requirements together with their corresponding system requirements and traceability links established by system engineers.}
The dataset includes the ``Lost Communication'' and ``Implausible Data'' DTC system requirements, which are among the most common and critical DTC categories. Each system requirement is linked to the corresponding stakeholder requirements through traceability links established by system engineers. Over time, system evolution across versions and the error-prone nature of manual traceability construction lead to a considerable number of invalid or missing traceability links. In total, the dataset comprises 1,320 stakeholder requirements connected to 48 system requirements through 2,132 existing traceability links.

\nff{To establish the ground truth, we followed a structured annotation protocol in collaboration with system engineers from our industry partner. First, two co-authors studied the industrial traceability guidelines, which define traceability relationships and validation criteria in the automotive DTC domain. They then conducted a calibration phase by independently annotating 100 randomly selected traceability links. Disagreements and ambiguous cases were discussed with system engineers to clarify the interpretations of the guidelines and identify recurring traceability variations observed in practice.}

\nff{
After calibration, the two annotators independently reviewed all 2,132 traceability links. Following the industrial guidelines, each stakeholder–system requirement pair was labeled as valid if the system requirement correctly covered the message, signal, fault condition, or diagnostic behavior specified in the stakeholder requirement, and as invalid otherwise. After this independent annotation phase, all conflicting annotations were jointly reviewed. For each disagreement, the annotators revisited the corresponding requirements and the industrial guidelines before reaching a final decision. The resulting labels were used as the ground truth for the experiments.}

\nff{To assess annotation reliability, we computed Cohen’s kappa before disagreement resolution, obtaining $\kappa \approx 0.962$, indicating almost perfect agreement. Ultimately, 1,913 links were confirmed as valid and 219 as invalid.}


\nff{}

In the experimental evaluation, due to the limited availability of manually validated traceability data, we adopt a leave-one-out cross-validation strategy~\cite {hastie2009elements}. Specifically, in each iteration, one traceability link (between stakeholder requirement and system requirement) is treated as the test instance, denoted as \(x_{\text{test}}\), while the remaining \(n-1\) links constitute a retrieval database, denoted as \(D_{\text{train}}\).

For each test instance, we apply a similarity-based retrieval method over \(D_{\text{train}}\) to identify representative in-context examples. In particular, the top-\(k\) most similar requirement pairs labeled as valid traceability links are selected as positive examples, while the top-\(k\) most similar requirement pairs labeled as invalid traceability links are selected as negative examples. These retrieved positive and negative examples are then used to guide the LLM in traceability validation and recovery. 

We acknowledge that, in practice, companies can obtain a larger volume of high-quality labeled data to enhance the retriever's database. As more traceability links are validated over time, the retrieval database can be continuously expanded, thereby providing richer, more representative in-context examples and potentially further improving the effectiveness of TVR.

In this study, the original dataset cannot be publicly disclosed due to data privacy constraints imposed by our industry partner. However, we provide fictitious, representative examples in the replication package~\cite{TVR2025} to illustrate the dataset structure and facilitate understanding.

\subsection{Evaluation Metrics}
\subsubsection{Traceability Validation}
To assess TVR's performance in automotive requirements traceability validation, we use standard metrics, including accuracy, precision, recall, F1 score, and Macro-F1 score. 
(see Formula~\ref{acc}, \ref{pre}, \ref{recall} and \ref{f}).
In our case, $TP$ is the number of traceability links that are correctly identified as valid. $FN$ is the number of traceability links that are incorrectly identified as invalid. $FP$ is the number of traceability links that are incorrectly identified as valid. $TN$ is the number of traceability links that are correctly identified as invalid.
\begin{equation}\label{acc}
{Accuracy = \frac{TP+TN}{TP+TN+FP+FN}}
\end{equation}

\begin{equation}\label{pre}
 {Precision = \frac{TP}{TP+FP}}
\end{equation}
 
\begin{equation}\label{recall}
 {Recall = \frac{TP}{TP+FN}}
\end{equation}
 
\begin{equation}\label{f}
{F1 = \frac{2\times Precision \times Recall}{Precision+Recall}}
\end{equation}

As shown in Formula~\ref{acc}, 
\textbf{Accuracy} is the proportion of correctly identified traceability links (both valid and invalid) out of all traceability links. \textbf{Precision} is the proportion of correctly identified valid links ($TP$) out of all links identified as valid by the model ($TP$ + $FP$). 
It measures the reliability of the model's predictions of valid links. 
\textbf{Recall} is the proportion of correctly identified valid links ($TP$) out of all actual valid links in the ground truth ($TP$ + $FN$). 
It measures the model's ability to find all the valid links. 
\textbf{F1 score} is the harmonic mean of precision and recall. 
It balances the trade-off between precision and recall, providing a single measure of the model's performance.
Given that our dataset is imbalanced, we also report \textbf{macro-F1}, the unweighted average of F1 scores across both classes. Macro-F1 treats each class equally and is therefore a more reliable metric for evaluating model performance on imbalanced datasets~\cite{schutze2008introduction}.

\subsubsection{Traceability Recovery} To evaluate TVR's performance on automotive requirements traceability recovery, two of our authors manually verified its \textit{Correctness}, and discussions were held to reach consensus when there was any disagreement. The evaluation metric is the proportion of correct traceability links, calculated as the number of links confirmed as correct through human verification divided by the total number of links identified by the model.
\begin{equation}\label{correctness}
    Correctness = \frac{Number\,of\,Correct\,Links\,Verified\,by\,Humans}{Total\,Number\,of\,Links\,Predicted\,by\,the\,Model}
\end{equation}

\subsection{Experiment Environment}
All experiments were conducted on a laptop provided by our industry partner, running Windows 10, equipped with an Intel Core i7-11850H CPU at 2.5 GHz and 32 GB of RAM. Due to data privacy concerns, we were limited to this single machine, which constrained our choice of text representation techniques. A temperature of 0 was applied to all LLMs to ensure consistent, reproducible outputs across experiments.

\subsection{Baselines}

To evaluate the effectiveness of TVR, we compare it with SOTA baselines. Since TVR targets the validation of traceability links between stakeholder and system requirements, we select baselines from two categories: (i) LLM-based validation approaches and (ii) retrieval-based approaches without validation.  

\noindent
\textbf{LLM-based Validation.}  
We adopt LiSSA~\cite{fuchss2025lissa}, a recent RAG-based approach for traceability link recovery (TLR). LiSSA first retrieves candidate \(\langle\)source, target\(\rangle\) pairs by computing cosine similarity scores between source and target requirements to identify the most similar pairs, and then validates the traceability link between them with GPT-4o under two prompting strategies: KISS (a simple Zero-Shot classification prompt~\cite{fuchss2025lissa}) and CoT. Since TVR performs only validation, we compare it against the validation component of LiSSA. To ensure fairness and protect data confidentiality (as our data cannot be exposed to OpenAI’s LLMs), we implemented LiSSA’s validation stage using the same LLM as TVR (i.e., Claude 3.5), while retaining the original KISS and CoT prompts.

\noindent
\textbf{Retrieval-based Approaches.}  
Following Fuchß et al.~\cite{fuchss2025lissa}, we also compare against retrieval-only baselines, which approximate the upper bound of retrieval performance by varying the similarity threshold. We implement two approaches following the implementation of Gao et al.~\cite{gao2022using}: (1) TF-IDF embeddings and (2) SentenceBERT embeddings~\cite{reimers2019sentence}, a widely used SOTA sentence-level embedding model. Cosine similarity is computed between stakeholder and system requirements, and a traceability link is identified if the similarity score exceeds the threshold. To approximate optimal performance, we vary the threshold from 0 to 1 in increments of 0.001 and report the maximum macro-F1 score, which serves as an upper bound for retrieval-based approaches.

\section{Experiment Results}\label{sec:results}

\subsection{Performance of SOTA LLMs on Requirements Traceability Validation (RQ1)}

\textbf{Approach.}
This RQ examines the performance of LLMs with Zero-Shot prompting for requirements traceability validation, focusing on models available to our industry partner, including Claude, Llama, Mistral, and Amazon Titan. Specifically, the Zero-Shot prompting strategy involves including the task description in the prompt without providing the model with any examples.

\noindent
\textbf{Results.} 
The experimental results are summarized in Table~\ref{tab:results}. Claude models generally outperform other model series, with Claude 2 (90.57\%), Claude Instant (84.43\%), and Claude 3.5 Sonnet (79.55\%) achieving the highest accuracy scores. In contrast, Llama 3 70B (10.19\%) and Mistral 8×7B (17.23\%) exhibit notably lower accuracy, indicating weaker generalization on this task. Claude 3.5 Sonnet achieved the highest Macro-F1 score of 58.75\%, surpassing all other models.

For valid pairs, most models exhibit high precision (above 90\%). Still, recall shows noticeable variation: Mistral Large 2402 achieves the highest precision (100\%) but has a recall of only 0.73\%, resulting in an F1 score of 1.44\%. This suggests the model is highly conservative, prioritizing precision over recall, but missing many valid pairs. Claude 2 achieves the best F1 score (95.04\%), with well-balanced precision (90.64\%) and recall (99.90\%), demonstrating strong predictive stability. Llama 3 70B and Mistral Large 2402 exhibit poor recall (0.95\% and 0.73\%, respectively), indicating difficulty capturing valid cases. Their F1 scores (1.88\% and 1.44\%) further confirm their limitations. Within the Claude series, Claude 2 outperforms other models, achieving the highest F1 score (95.04\%), while Claude Instant follows with 91.51\%.

However, detecting invalid pairs proves more challenging, as evidenced by the lower F1 scores across all models. Claude 2 has the highest invalid precision (71.43\%) but suffers from extremely low recall (2.45\%), resulting in an F1 score of only 4.74\%, indicating it misses most invalid pairs. Mistral Large 2402 achieves the highest recall for invalid cases (100.00\%) but at the cost of low precision (9.63\%), resulting in an F1 score of 17.57\%. This suggests it identifies all invalid pairs, but at the expense of many false positives. Among weaker performers, Claude Instant has an invalid recall of only 4.90\% and an F1 score of 5.68\%. Llama 3 70B achieves high recall (96.53\%) but low precision (9.44\%), resulting in an F1 score of 17.20\%.

The reasons for the models' poor performance include: (1) Even when explicitly instructed to focus on determining whether a message or signal from the stakeholder requirement is covered by the linked system requirement, without domain knowledge, LLMs still struggle to accurately understand and identify the key relevant message or signal in the requirements, but rather focus on the wrong aspects, leading to incorrect predictions. (2) LLMs tend to check for content consistency between requirements and often misjudge the two requirement descriptions as inconsistent due to differences in wording or structure. 
Since invalid pairs are few, the small denominators in score calculations, such as precision and recall, can lead to unreliable scores. The Macro-F1 scores are also low across all LLMs (below 60\%).


\answer{
\textit{\textbf{Answering RQ1:} Results suggest unsatisfactory accuracy for all models with Zero-Shot prompting for requirements traceability validation, especially for invalid links, 
thus underscoring the necessity of employing more advanced prompt engineering strategies.}}

\begin{table}[htbp]
\caption{Experimental Results of LLMs with Different Prompting Strategies and our RAG-based TVR Approach (\%).}
\label{tab:results}
\small
\begin{tabular}{|p{1.2cm}|p{2.8cm}|p{0.75cm}|p{0.75cm}|p{0.75cm}|p{0.75cm}|p{0.75cm}|p{0.75cm}|p{0.75cm}|p{0.75cm}|}
\toprule
\multirow{2}{*}{\textbf{Prompt}} &\multirow{2}{*}{\textbf{Model}} & \multirow{2}{*}{\textbf{Acc}} & \multicolumn{3}{c|}{\textbf{Valid}}  & \multicolumn{3}{c|}{\textbf{Invalid}} & \textbf{macro-} \\  \cmidrule(lr){4-6}  \cmidrule(lr){7-9}
&  & & Pre & Recall & F1 & Pre & Recall & F1 & \textbf{F1} \\ \midrule 
\multirow{16}{*}{Zero-Shot} 
& Claude 3.5 Sonnet               & 79.55       & 93.42  & 83.25 & 88.04 & 21.98 & 44.61  & \cellcolor[HTML]{EFEFEF}  29.45 & \cellcolor[HTML]{EFEFEF} 58.75 \\
& Claude 3 Sonnet                 & 32.08       & 96.51  & 25.83 & 40.75 & 11.51 & 91.18  & 20.44 & 30.60 \\
& Claude 2                        & \cellcolor[HTML]{EFEFEF} 90.57       & 90.64  & \cellcolor[HTML]{EFEFEF} 99.90 & \cellcolor[HTML]{EFEFEF} 95.04 & \cellcolor[HTML]{EFEFEF} 71.43 & 2.45   & 4.74 & 49.89 \\
& Claude Instant                  & 84.43       & 90.22  & 92.84 & 91.51 & 6.76  & 4.90   & 5.68 & 48.60 \\
& Claude 3 Haiku                  & 49.72       & 91.00  & 49.27 & 63.93 & 10.11 & 53.92  & 17.03 & 40.48 \\ 
& Llama 3 8B                      & 72.70       & 91.09  & 77.39 & 83.68 & 11.74 & 28.43  & 16.62 & 50.15\\
& Llama 3 70B                     & 10.19       & 72.00  & 0.95  & 1.88  & 9.44  & 96.53  & 17.20 & 9.54 \\ 
& Mistral 7B                      & 57.09       & 91.96  & 57.45 & 70.72 & 12.08 & 53.77  & 19.72 & 45.22 \\
& Mixtral 8x7B                    & 17.23       & 89.76  & 9.62  & 17.38 & 9.43  & 89.55  & 17.07 & 17.23\\
& Mistral Large 2402              & 10.23       & \cellcolor[HTML]{EFEFEF}100.00 & 0.73  & 1.44  & 9.63  & \cellcolor[HTML]{EFEFEF}100.00 & 17.57 &9.51\\ 
& Titan Text Premier       & 20.68       & 92.17  & 13.43 & 23.45 & 9.83  & 89.22  & 17.71 & 20.58\\
& Titan Text Express       & 49.81       & 90.17  & 49.95 & 64.29 & 9.30  & 48.53  & 15.62 & 39.96 \\
& Titan Text Lite          & 70.83       & 91.49  & 74.69 & 82.24 & 12.54 & 34.31  & 18.37 & 50.31 \\ \midrule
\multirow{16}{*}{CoT} 
& Claude 3.5 Sonnet  & 76.31       & 98.70\,\textuparrow  & 74.79 & 85.10 & \cellcolor[HTML]{EFEFEF}27.57\,\textuparrow & 90.69\,\textuparrow  & \cellcolor[HTML]{EFEFEF} 42.29\,\textuparrow & \cellcolor[HTML]{EFEFEF} 63.7\,\textuparrow \\
& Claude 3 Sonnet                 & 22.92       & 95.24  & 15.60 & 26.81 & 10.33 & 92.57\,\textuparrow  & 18.59 & 22.7 \\
& Claude 2                        & 66.47       & 94.39\,\textuparrow  & 66.79 & 78.22 & 17.24 & 63.54\,\textuparrow  & 27.12\,\textuparrow & 52.67$\uparrow$ \\
& Claude Instant                  & 86.16\,\textuparrow       & 90.36\,\textuparrow  & \cellcolor[HTML]{EFEFEF} 94.81\,\textuparrow & 92.53\,\textuparrow & 8.26\,\textuparrow  & 4.41    & 5.75\,\textuparrow & 49.14$\uparrow$ \\
& Claude 3 Haiku                  & 45.55\,\textuparrow       & 94.97\,\textuparrow  & 41.62 & 57.88 & 13.44\,\textuparrow & 80.42\,\textuparrow  & 23.03\,\textuparrow & 40.46 \\ 
& Llama 3 8B                      & \cellcolor[HTML]{EFEFEF} 86.25\,\textuparrow      & 90.56  & 94.73\,\textuparrow & \cellcolor[HTML]{EFEFEF} 92.59\,\textuparrow & 6.00  & 3.30   & 4.26 & 48.43 \\
& Llama 3 70B                     & 35.32\,\textuparrow       & 98.17\,\textuparrow  & 28.94\,\textuparrow & 44.70\,\textuparrow & 12.51\,\textuparrow & 94.97  & 22.11\,\textuparrow & 33.41$\uparrow$ \\ 
& Mistral 7B                      & 73.15\,\textuparrow       & 92.35\,\textuparrow  & 76.63\,\textuparrow & 83.76\,\textuparrow & 15.65\,\textuparrow & 40.59  & 22.59\,\textuparrow & 53.18$\uparrow$ \\
& Mixtral 8x7B                    & 14.53       & 96.43\,\textuparrow  & 5.80  & 10.94 & 9.82\,\textuparrow & 97.95\,\textuparrow  & 17.84\,\textuparrow & 14.39 \\
& Mistral Large 2402              & 10.27\,\textuparrow       & \cellcolor[HTML]{EFEFEF} 100.00 & 0.78\,\textuparrow & 1.54\,\textuparrow  & 9.64\,\textuparrow  & \cellcolor[HTML]{EFEFEF} 100.00 & 17.58\,\textuparrow & 9.56$\uparrow$ \\ 
& Titan Text Premier       & 25.70\,\textuparrow      & 92.57\,\textuparrow  & 19.40\,\textuparrow & 32.08\,\textuparrow & 10.07\,\textuparrow & 85.29  & 18.01\,\textuparrow & 25.05$\uparrow$ \\
& Titan Text Express       & 54.41\,\textuparrow      & 90.78\,\textuparrow  & 55.19\,\textuparrow & 68.65\,\textuparrow & 10.00\,\textuparrow & 47.06  & 16.49\,\textuparrow & 42.57$\uparrow$ \\
& Titan Text Lite          & 69.93\,\textuparrow      & 90.70\,\textuparrow  & 74.38 & 81.73 & 10.34 & 27.94  & 15.10 & 48.42 \\ \midrule
\multirow{11}{*}{Few-Shot}
& Claude 3.5 Sonnet               & \cellcolor[HTML]{EFEFEF} 97.47\,\textuparrow	    & 98.70 & 98.50\,\textuparrow	 & \cellcolor[HTML]{EFEFEF} 98.60\,\textuparrow	& \cellcolor[HTML]{EFEFEF}86.06\,\textuparrow	& 87.75	& \cellcolor[HTML]{EFEFEF} 86.89\,\textuparrow & \cellcolor[HTML]{EFEFEF} 92.75$\uparrow$ \\
& Claude 3 Sonnet                 & 90.81\,\textuparrow	    & 94.00 & 95.95\,\textuparrow  & 94.97\,\textuparrow    & 52.44\,\textuparrow & 42.16 & 46.74\,\textuparrow & 70.86\,\textuparrow \\
& Claude 2                        & 41.44       & 97.88\,\textuparrow & 36.01  & 52.66    & 13.29 & 92.65\,\textuparrow & 23.25 & 37.96 \\
& Claude Instant                  & 90.95\,\textuparrow       & 96.82\,\textuparrow  & 93.05 & 94.90\,\textuparrow & 51.97\,\textuparrow & 71.08\,\textuparrow  & 60.04\,\textuparrow & 77.47$\uparrow$ \\
& Claude 3 Haiku                  & 92.73\,\textuparrow       & 93.65  & 98.65\,\textuparrow & 96.08\,\textuparrow & 74.26\,\textuparrow & 36.76  & 49.18\,\textuparrow & 72.63$\uparrow$ \\
& Mistral 7B                      & 90.01\,\textuparrow       & 90.62  & \cellcolor[HTML]{EFEFEF} 99.22\,\textuparrow & 94.73\,\textuparrow & 28.57\,\textuparrow & 2.94   & 5.33  &50.03 \\
& Mixtral 8x7B                    & 30.66\,\textuparrow       & \cellcolor[HTML]{EFEFEF} 99.13\,\textuparrow  & 23.56\,\textuparrow & 38.07\,\textuparrow & 11.90\,\textuparrow & \cellcolor[HTML]{EFEFEF} 98.03\,\textuparrow  & 21.23\,\textuparrow & 29.65$\uparrow$ \\
& Mistral Large 2402              & 61.35\,\textuparrow       & 98.94  & 57.88\,\textuparrow & 73.04\,\textuparrow & 19.12\,\textuparrow & 94.12  & 31.79\,\textuparrow & 52.42$\uparrow$ \\
\midrule
\multirow{1}{*}{Self-Cons.}
& Claude 3.5 Sonnet               & 97.61\,\textuparrow      & 98.4   & 98.96\,\textuparrow  & 98.68\,\textuparrow  & 89.64\,\textuparrow & 84.8  & 87.15\,\textuparrow & 92.92$\uparrow$  \\ \midrule
\multirow{1}{*}{TVR} 
& Claude 3.5 Sonnet               & \textbf{98.87}       & \textbf{99.28}  & \textbf{99.48}   & \textbf{99.38}   & \textbf{95.00}  & \textbf{93.14}  & \textbf{94.06}  & \textbf{96.72} \\ 
\bottomrule
\multicolumn{10}{c}{\textbf{Baselines}} \\ \bottomrule
\multirow{1}{*}{LiSSA-KISS}  & \multirow{2}{*}{Claude 3.5 Sonnet} & 74.77 & 91.54 & 79.19 & 84.92 & 16.56 & 30.07 & 22.70  & 53.81 \\ 
\multirow{1}{*}{LiSSA-CoT} & & 70.73 & 92.10 & 73.71 & 81.88 & 16.31 & 44.75 & 23.90 & 52.89 \\ \hline
\multirow{2}{*}{Retrieval} & TF-IDF (0.003)& 78.94 & 89.61 & 86.57 & 88.06 & 9.51 & 12.33 & 10.74 & 49.40 \\
 & SBERT (0.287) & 86.30 & 91.12 & 93.88 & 92.48 & 27.33 & 20.09 & 23.16& 57.82 \\ \bottomrule
\end{tabular}
\end{table}

\subsection{Best Prompt Strategy for Requirements Traceability Validation (RQ2)}

\noindent
\textbf{Approach.}
This RQ explores the performance of different prompt engineering strategies (i.e., CoT and Few-Shot) and our RAG-based TVR approach for the requirements traceability validation task.

For Few-Shot prompting, we consider all four distinct variations of stakeholder requirements presented in our context, each corresponding to either \textit{mature} or \textit{demature} conditions of the linked system requirements. Additionally, the traceability link between each stakeholder requirement and its corresponding system requirement can be valid or invalid. Given these factors, our dataset contains 16 possible combinations.
To maximize the effectiveness of LLMs, we randomly selected one example from each combination, resulting in a total of 16 examples for the prompt. However, the available Llama series models (Llama 3 8B and Llama 3 70B) and Amazon Titan series models (Text Premier, Text Express, and Text Lite) have a limited token capacity (8K), which is insufficient for a 16-shot prompt. Consequently, we excluded these five models from our Few-Shot prompt experiments, leaving eight models for requirements traceability validation. For these models, we include all 16 examples in the prompt.

Based on the results shown in Table~\ref{tab:results}, we select the best-performing model (i.e., Claude 3.5 Sonnet) for Few-Shot prompting and then evaluate self-consistency on this model. Specifically, we run the model 10 times and use majority voting, selecting the most frequent output as the model's final output~\cite{wang2022self}.

For RAG-based TVR, we also use Claude 3.5 Sonnet and then use the approach described in Section~\ref{sec:approach} to retrieve the \( k \times 2 \) most similar examples as part of the prompt. Due to the limited amount of manually labeled ground-truth data, we employ leave-one-out cross-validation~\cite{hastie2009elements} in our experiments to make efficient use of the data and ensure realistic estimates. In each iteration, one traceability link is used as a test, while the remaining \( n-1 \) links serve as the retrieval database. For the retriever component, we experimentally compare two similarity measures: cosine similarity and Euclidean distance, and evaluate the impact of different values of \textit{k}, ranging from 1 to 8.

\noindent
\textbf{Results.} The experiment results of different models with various prompt engineering strategies, as well as the RAG-based TVR approach, are reported in Table~\ref{tab:results}. We use ``$\uparrow$'' to highlight improved scores compared to the previous prompt strategy in the table. The results of the four baselines are also reported at the bottom of Table~\ref{tab:results}.

The experimental results indicate that incorporating CoT reasoning enhances recall across most models, both for valid and invalid cases. Claude 3.5 Sonnet, Claude 2, and Claude 3 Haiku exhibit substantial improvements in invalid recall, leading to higher F1 scores. Meanwhile, Llama 3 8B, Llama 3 70B, Mistral 7B, and Amazon Titan Text Premier show notable gains in valid F1 scores. However, for Claude 3 Sonnet, Llama 3 8B, and Amazon Titan Text Lite, invalid F1 scores decreased due to a drop in precision when detecting invalid pairs.

Comparing the results of Few-Shot prompting with CoT prompting, we can observe that adding examples to the prompt effectively improves the F1 score for most models in both categories (valid and invalid pairs), except for Claude 2 and Mistral 7B. Apart from Claude 2, all models show improved precision in the invalid category. This demonstrates that including examples in the prompt helps models better distinguish valid from invalid requirement pairs, achieving higher precision for invalid pairs. However, the downside of adding too many examples is that it increases the prompt length, which may exceed the token limit of some models.

For Few-Shot prompting, the best-performing model is Claude 3.5 Sonnet. It outperforms all other models in terms of average performance across both categories. It achieves over 98\% in precision, recall, and F1 score for the valid category, and over 86\% for the invalid category. The macro-F1 score of 92.75\% still surpasses all other models. 
However, applying the self-consistency approach to Claude 3.5 Sonnet does not improve its overall performance, as the Macro-F1 scores are almost identical (92.75\% for Few-Shot and 92.92\% for Self-Consistency).

\begin{figure}[!htbp]
\centering
\includegraphics[width=0.6\linewidth]{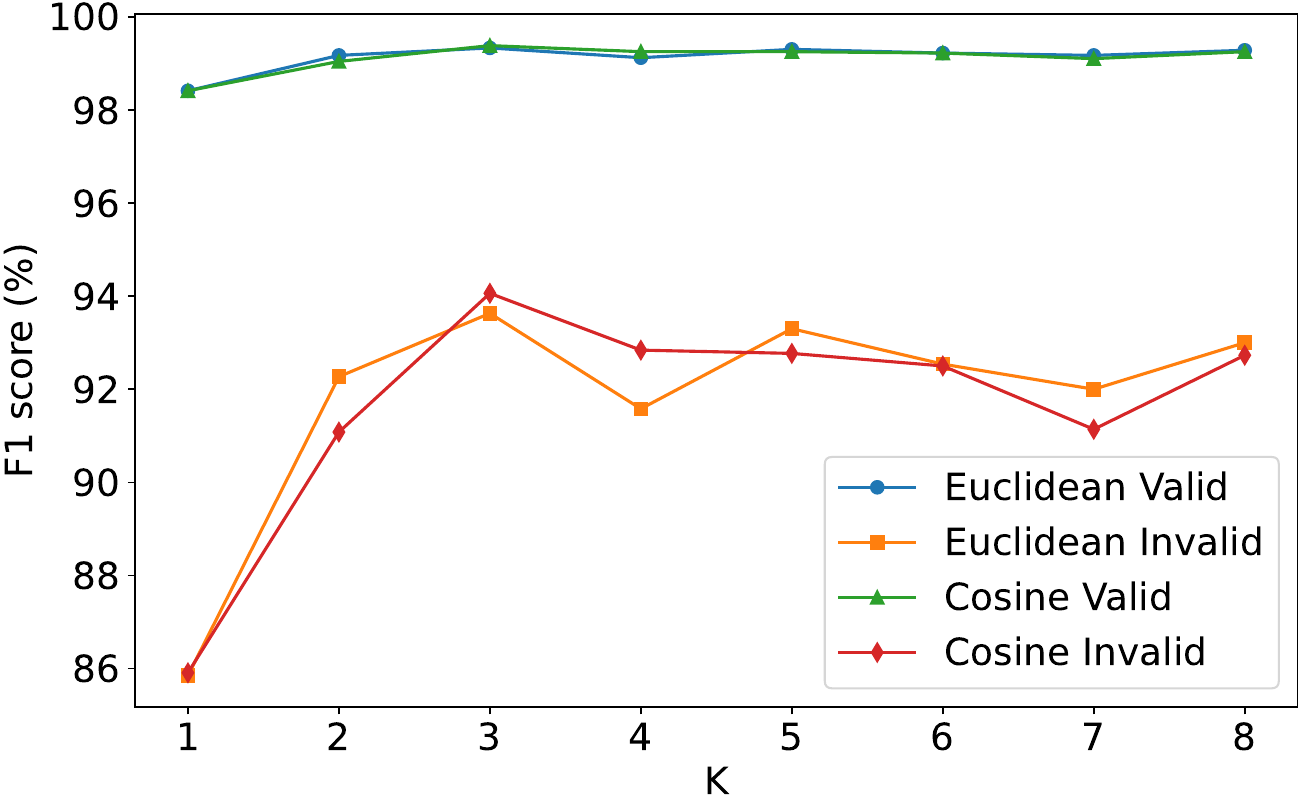}
\caption{F1 Score Comparison for Euclidean and Cosine Similarity Across Different Values of \textit{K}.}
\label{fig:k}
\end{figure}

For RAG, we first compare TVR performance across different \textit{K}s (i.e., the number of examples in the prompt) and distance functions. As shown in Figure~\ref{fig:k}, the optimal configuration of TVR uses cosine similarity with \textit{K}=3, which reaches the highest F1 score for both valid and invalid pairs. TVR with Claude 3.5 Sonnet achieved the highest accuracy (98.87\%) and a Macro-F1 score of 96.72\% across all configurations. In addition, Fisher’s exact test results show that its accuracy is higher than that of other configurations, with $p = 0.0011$ and $p = 0.0004$ when compared with self-consistency and few-shot learning, respectively, both of which are below the 0.05 significance level. This test compares proportions of correctly predicted valid and invalid pairs. 

When compared with the four baselines, TVR consistently and significantly outperforms all of them, as confirmed by Fisher’s exact test results ($p = 5.83 \times 10^{-174}$, $2.03 \times 10^{-143}$, $6.05 \times 10^{-64}$, and $2.74 \times 10^{-113}$ for LiSSA-CoT, LiSSA-KISS, SentenceBERT, and TF-IDF, respectively). LiSSA achieves Macro-F1 scores of 53.81\% and 52.89\% with the KISS and CoT prompts, respectively, while the retrieval-based baselines, TF-IDF and SentenceBERT, reach upper bounds of 49.40\% and 57.82\% in terms of Macro-F1 scores at similarity thresholds of 0.003 and 0.287. This performance gap with TVR  is explainable: LiSSA’s prompts provide no examples, making it difficult for LLMs to accurately define “traceability,” particularly in the automotive domain. For the retrieval-based baselines, DTC requirements involve domain-specific terminology, including signals and messages; thus, purely lexical or semantic similarity often fails to capture the true traceability links. These results highlight that in industrial settings, dedicated approaches such as TVR are necessary, whereas general-purpose methods may be insufficient.

\nff{To further investigate the relationship between TVR and retrieval-based approaches, we analyzed the misclassified cases of TVR and the strongest retrieval baseline (SentenceBERT). Interestingly, we found that the two approaches exhibited completely disjoint error sets. Specifically, all 24 samples misclassified by TVR were correctly classified by SentenceBERT, whereas all 292 samples misclassified by SentenceBERT were correctly classified by TVR~\footnote{Notably, the SentenceBERT results correspond to an optimistic upper bound obtained by exhaustively selecting the optimal similarity threshold over the entire dataset. Therefore, the practical performance of retrieval-based approaches is expected to be lower than reported here.}. This observation indicates that the two approaches rely on fundamentally different decision mechanisms and therefore exhibit orthogonal error patterns. Although retrieval-based methods achieve substantially lower overall performance, they can provide complementary evidence in rare cases where TVR fails. These findings suggest that retrieval-based similarity and LLM-based reasoning capture different aspects of traceability evidence, motivating future hybrid approaches that combine the strengths of both techniques.}

\nff{To better understand the limitations of TVR, we manually analyzed the misclassified cases and observed two dominant error patterns. Most false negatives occurred because the LLM occasionally considered stakeholder requirement components beyond the target message or signal, such as trigger conditions, mitigation actions, validation rules, or reference documents. Although our prompt explicitly instructs the model to focus only on the input message or signal, the model sometimes judged a traceability link as invalid because these additional components were not explicitly covered by the system requirement. Most false positives occurred when messages appearing in trigger conditions were mistakenly treated as the target message for traceability validation. In such cases, the LLM incorrectly inferred that the system requirement covered the stakeholder requirement because the trigger-condition message appeared in both requirements. These findings suggest that the primary limitation of TVR is the occasional inability of LLMs to consistently adhere to the intended validation criterion. 
Future work could explore more constrained prompting strategies or explicitly identifying different requirement components (e.g., trigger condition, input message, and mitigation actions) before validation to further reduce such errors.
Another interesting direction is to integrate retrieval-based similarity with LLM-based reasoning. Although retrieval alone performs substantially worse than TVR, our analysis shows that the two approaches exhibit complementary error patterns, suggesting that hybrid approaches may further improve traceability validation performance.
}


In conclusion, TVR using Claude 3.5 Sonnet with RAG achieved the best performance in requirements traceability validation, with a maximum accuracy of 98.87\% and Macro-F1 of 96.72\% across all configurations. It also demonstrated improvement over baselines, highlighting the effectiveness of retrieving similar examples and incorporating them into the prompt, thereby helping the LLMs better understand the requirement pair to be validated. 


\answer{
\textit{\textbf{Answering RQ2:} Claude 3.5 Sonnet with RAG achieved the best performance in requirements traceability validation when using cosine similarity with \textit{K} set to 3. It outperforms baselines, making it a viable and highly accurate solution for traceability validation in practice.}}
 
\subsection{Robustness to Unseen Requirement Variations (RQ3)}

\textbf{Approach.} This RQ evaluates the robustness of TVR to unseen variations of stakeholder requirements, a crucial aspect in practice. To ensure a comprehensive evaluation, we employ cross-validation by categorizing all requirement pairs by their variations and evaluating across the four categories in our dataset. Specifically, for each requirement pair to be validated, our retriever component first identifies its variation category. Then it retrieves the most similar examples from the remaining three variation categories, thereby emulating situations in which new variations are encountered. 

\noindent
\textbf{Results.} Table~\ref{tab:robustness} presents the TVR robustness evaluation results. The Macro-F1 scores, ranging from 87.02\% to 94.92\%, demonstrate TVR’s strong generalization ability across variation categories.

\begin{table}[!htbp]
\caption{Robustness Evaluation Results.}
\begin{tabular}{cc|ccc|ccc|c}
\hline
    & \multirow{2}{*}{Acc(\%)} & \multicolumn{3}{c|}{Valid} & \multicolumn{3}{c|}{Invalid} & Macro- \\
    &                          & Pre(\%)  & Recall(\%) & F(\%)     & Pre(\%)  & Recall(\%)  & F(\%)  & F1(\%)    \\\hline
All & 97.13                    & 97.40     & 99.48  & 98.43 & 93.83    & 74.88   & 83.29  & 90.86 \\
V1  & 93.96 &	94.37 &	98.75 &	96.51 &	90.79 &	67.65 &	77.53 & 87.02 \\
V2  & 98.78	& 98.81 &	99.89  &	99.35 &	98.28 &	83.82 &	90.48 & 94.92 \\
V3  & 98.44 & 98.59 &	99.76 &	99.18 &	95.24 &	76.92 &	85.11 & 92.15\\
V4  & 92.59 &	95.00 &	95.00 &	95.00 &	85.71 &	85.71 &	85.71 & 90.36\\ \hline
\end{tabular}
\label{tab:robustness}

\end{table}

Among variation categories, V4 shows the lowest accuracy (92.59\%). V1 shows the lowest Macro-F1 score (87.02\%), with a noticeable drop in recall for invalid pairs (67.65\%), suggesting frequent misclassifications of invalid pairs as valid. In contrast, results for V2 and V3 are similar, with high accuracy (98.78\% and 98.44\%) and Macro-F1 scores (94.92\% and 92.15\%), demonstrating strong adaptability.  

These results may be attributed to two factors: (1) Variability in category differences affects the quality of retrieved examples in guiding TVR. Specifically, V1 belongs to the ``Lost Communication'' type of DTCs, whereas the remaining three variations fall under the ``Implausible Data'' category. This discrepancy likely contributed to TVR’s poorer cross-validation performance on V1. (2) Certain variations exhibit greater complexity, making them more challenging for the LLM to understand and predict accurately. Although V2, V3, and V4 all belong to the ``Implausible Data'' category, our analysis revealed that V4 is more complex than other variations, 
containing more messages and conditions to be verified. Moreover, the V4 template is not strictly fixed, resulting in slightly lower cross-validation accuracy. 


\answer{
\textit{\textbf{Answering RQ3:} TVR achieved an overall accuracy of 97.13\% and Macro-F1 score of 90.86\% on robustness evaluation, indicating its strong generalization ability across unseen variation categories, with more challenges for certain variation categories that are more distinct and invalid pairs.}}

\subsection{Performance on Requirements Traceability Links Recovery (RQ4)}
\textbf{Approach.} As explained in Section~\ref{sec:retrieving}, we employ a three-step preprocessing process to reduce the number of requirement pairs we consider that could have a missing link. For the remaining pairs, TVR predicts whether a valid traceability link exists between them. 
We then calculated the \textit{Correctness} of predicted links.

\noindent
\textbf{Results.} The three stages of preprocessing yielded, in turn, 30,598, 14,494, and 1,919 requirement pairs. This demonstrates the effectiveness of our preprocessing, which reduces the number of infeasible pairs. Then, 502 pairs out of 1,919 pairs are predicted to have valid traceability links. After manually verifying all 502 pairs ($\kappa \approx 0.92$), we obtained a \textit{Correctness} of 85.50\%, suggesting that most retrieved links are correct, with a limited number of false positives.

We conducted an error analysis and found that all 73 prediction errors stemmed from the same issue. In the stakeholder requirement, TVR first verifies whether the trigger condition is set to ``RUN'', then checks the input message to determine whether the DTC is being set or cleared. Since a trigger condition is also a message, if the corresponding system requirement includes the same message, the model may incorrectly interpret this as a traceability link, whereas it should actually check for the input message. To enhance retrieval accuracy, future approaches can be improved by identifying and excluding the trigger condition, retaining only content related to the input message.

\answer{
\textit{\textbf{Answering RQ4:} After effectively reducing the number of requirement pairs being considered with our three-step filtering approach, TVR achieves an 85.50\% correctness in recovering missing links, thus making it applicable with a manageable number of false positives. Based on an error analysis, we conclude that additional preprocessing of messages and signals may further improve accuracy.}}

\section{Discussion}~\label{sec:discussion}
\subsection{Industry Feedback on TVR}
\nff{To gain additional industrial insights, we presented TVR and its experimental results to practitioners from our industrial partner during a project meeting. The participants included two engineers who are actively involved in requirements engineering and traceability-related activities. During the meeting, we presented the quantitative evaluation results for TVR, including its precision, recall, and F1-score. We also demonstrated representative examples of TVR's reasoning process and explanations. The feedback reported in this section was collected through the subsequent discussion session.}

Overall, the participants expressed positive opinions regarding TVR. They highlighted its high accuracy and reliability, recognizing its potential to support consistency checks for traceability in industrial practice. Participants also appreciated the high level of automation provided by TVR, which could significantly reduce manual effort in traceability verification, leading to substantial time and cost savings.


The participants also valued the interpretability of TVR's decisions. To illustrate this capability, we demonstrated TVR's reasoning process by removing the instruction ``\textit{and only respond with either `Yes' or `No'}'' from our prompt. In its response, TVR first identifies the key message and signal in the stakeholder requirement. Then, it searches for the relevant message or signal in the linked system requirement and checks whether the system requirement addresses the message or signal in the stakeholder requirement, taking consistent action accordingly. For an invalid pair, TVR provides an example explanation as follows:

\noindent
\textit{``\textbf{Step 1}: Identify the key message/signal in the stakeholder requirement. The key message is MESSAGE\_1. \textbf{Step 2}: Search for this message in the system requirement. After carefully examining the system requirement, I don't see any explicit mention of MESSAGE\_1 or any checks related to it. \textbf{Step 3}: Determine if the message is covered. Since there is no explicit mention or check for MESSAGE\_1, we can conclude that this specific message is not covered. \textbf{Step 4}: Conclusion. The response is: No.''} 


According to the participants, such explanations can help engineers quickly understand why a traceability link is classified as valid or invalid, thereby facilitating manual verification and increasing confidence in automated traceability decisions.

Looking ahead, industry professionals identified two key points for improvement: Further minimizing false positives to further refine precision and enhancing scalability to better adapt to large-scale industrial datasets with diverse requirements. These insights will inform future refinements, ensuring the approach remains both effective and practical for real-world adoption.

\subsection{Practical Application of TVR}
TVR is designed to automate and enhance the accuracy of traceability link validation and recovery for DTC requirements in automotive systems. To apply TVR, practitioners only need to prepare three key components:  
(1) The input pairs to be verified, i.e., stakeholder requirements and system requirements. 
(2) The retrieval database, i.e., a collection of requirement pairs with labels (valid or invalid) previously validated by system engineers.
(3) The prompt specifying the task description and instructions.
Once these inputs are provided, TVR generates a prediction result with an accompanying explanation, as instructed by the prompt, thereby improving transparency and interpretability.  

TVR is particularly well-suited for iterative software development in industrial settings. As the retrieval database expands with more human-confirmed data, the retriever can extract more relevant examples, enhancing the LLM’s ability to make more accurate and reliable assessments over time.  

When applying TVR to different datasets (beyond DTC requirements) or to various software artifacts, the prompt instructions must be adjusted to ensure the task description closely aligns with the specific problem at hand. Additionally, selecting the appropriate number of examples and the similarity measure requires empirical experimentation with the specific data at hand to determine the optimal configuration. A priori, the basic principles of TVR can also be adapted to other domains and types of requirements. 
Moreover, the comparison with baselines demonstrates that a general prompt may not perform well in specific real-world scenarios.

In summary, TVR demonstrates excellent performance on industrial data, offering high accuracy, strong robustness, ease of use, and broad applicability.

\section{Threats to Validity}\label{sec:threats}

\noindent

\noindent

\noindent
\textbf{External Validity.}
TVR is specifically designed for DTC requirements critical to the automotive domain, using a dataset from an industry partner. 
Since TVR relies on a retrieval-based approach to obtain prompt examples, its accuracy in practical applications may be affected by the quality of the retrieval database. 
\nff{Furthermore, the dataset used in this study was collected from a single industrial automotive project. Although this dataset covers multiple DTC categories, four distinct variations in stakeholder requirements, and numerous domain-specific messages and signals, the observed results may not generalize to other projects, product lines, organizations, or domains with different requirements engineering practices and conventions. Future studies should evaluate TVR on additional industrial projects and product lines to further assess its generalizability.}
Nevertheless, the general principles of TVR should be widely applicable to many types of requirements in domains where traceability between high-level stakeholder requirements and system requirements is important.

\noindent
\textbf{Internal Validity.}
Co-authors annotated the dataset under the guidance of system engineers. However, manual annotation is inherently prone to errors, such as misinterpretation of requirements or inconsistencies in labeling. These annotation errors could affect the evaluation of TVR’s performance. To mitigate this threat, two of our authors independently labeled the dataset and resolved any conflicts through discussion.
LLMs may produce different outputs for the same input due to their stochastic nature. We mitigated this threat by setting the temperature to 0 to ensure consistent and deterministic outputs. Another internal threat arises from the model's robustness evaluation, which is influenced by the degree of variation across categories. To mitigate this threat, we used cross-validation to evaluate the model's robustness across all four categories. For traceability link recovery, rather than relying on similarity-based methods, we adopted a three-step rule-based preprocessing strategy to filter requirement pairs. These manually defined rules were designed to ensure that no true traceability links were inadvertently removed. Their applicability to other datasets or domains may be limited, raising concerns about external validity.
Lastly, the set of models we used was constrained by the industry partner’s data access policy. In future work, we plan to evaluate TVR on open-source models and datasets, as well as more recent models. Nonetheless, the results demonstrate that TVR is highly effective (96.72\% Macro-F1 score), outperforming existing baselines. Results could be improved with other models, but our conclusions would remain unaffected. Though agent-based methods could be considered, the potential for improvement is small, and such techniques would likely be less efficient in terms of time and cost. 

\section{Related Work}\label{sec:relatedwork}
Traceability, defined as ``the ability to describe and follow the life of an artifact developed during the software lifecycle in both forward and backward directions'' ~\cite{lucia2007recovering}, plays a crucial role in requirements engineering and software engineering in general. To date, the most extensively studied challenges are TLR and traceability maintenance~\cite{guo2024natural}. Researchers have proposed various approaches to support software traceability between different artifacts, including requirements-to-code~\cite{gao2022using, hey2021improving, panichella2013and, kuang2015can}, document-to-code~\cite{antoniol2002recovering, marcus2003recovering, keim2024recovering}, and document-to-model~\cite{keim2023detecting, cleland2005utilizing}.  

Early automated traceability techniques relied on classical NLP approaches, primarily leveraging information retrieval techniques~\cite{guo2024natural}. These approaches establish potential traceability links by computing textual similarity between artifacts using models such as the vector space model~\cite{antoniol2002recovering, mahmoud2015information}, latent semantic indexing (LSI)~\cite{marcus2003recovering}, latent Dirichlet allocation (LDA)~\cite{asuncion2010software}, and hybrid approaches~\cite{gethers2011integrating, moran2020improving}. However, these traditional models struggle to capture deep semantics. Recent approaches have addressed this limitation by incorporating word embeddings and deep neural networks~\cite{guo2017semantically, wang2018enhancing, zhao2017improved, hey2021improving}. Additionally, some studies have explored active learning~\cite{mills2019tracing} and self-attention mechanisms~\cite{zhang2021recovering} for improving TLR.

With the advent of LLMs such as GPT~\cite{openai2023chatgpt, openai2023gpt4}, Llama~\cite{llama}, and Claude~\cite{claude}, new opportunities have emerged for automating traceability while addressing the limitations of previous approaches. Hey et al.~\cite{hey2021improving} proposed the FTLR approach, which integrates word embedding similarity and Word Mover's Distance~\cite{kusner2015word} to generate requirements-to-code trace links. Hey et al.~\cite{hey2024requirements} further enhance accuracy by incorporating a NoRBERT classifier to filter irrelevant requirement segments. Rodriguez et al.~\cite{rodriguez2023prompts} investigated the use of Claude to directly perform TLR through natural language prompting. Their study demonstrated that generative models can not only recover trace links but also provide explanations for identified links. Fuchss et al.~\cite{fuchss2025lissa} recently introduced LiSSA, a RAG-enhanced LLM approach for TLR across requirements-to-code, documentation-to-code, and architecture documentation-to-models. LiSSA first retrieves relevant elements based on similarity values between embeddings, then uses an LLM to validate traceability links. 

Similarly, in a recent study, Hey et al.~\cite{hey2025requirements} use GPT-4o to automate inter-requirement TLR between high-level and low-level requirements. Their approach first retrieves similar requirements, then employs Zero-Shot and CoT prompting to validate the traceability links. However, existing approaches achieve limited accuracy, as they rely solely on simple Zero-Shot or CoT prompts combined with generic queries (e.g., ``Is there a traceability link between...''). These approaches are inadequate for precisely validating traceability links between the DTC system and stakeholder requirements, as evidenced by the results in Table~\ref{tab:results}. DTC requirements often vary slightly in the character strings of messages and signals. Consequently, such general-purpose approaches struggle to capture the subtle nuances of domain-specific requirements.

In contrast to prior studies, our research focuses on validating traceability links as well as recovering missing links, addressing specific industry needs in the automotive sector: 1) Industry practitioners require a mechanism to validate the correctness of traceability links between stakeholder requirements and system requirements established by system engineers. 2) Both stakeholder and system requirements exhibit variations in the way they are expressed, with the possibility of encountering unseen variations in the future. 3) Both stakeholder and system requirements follow loose templates, with small differences in the way message and signal values are expressed, thus rendering information retrieval-based approaches based on term frequency and word overlapping ineffective. To address these challenges, we frame the validation task as a binary classification problem in which LLMs determine whether existing traceability links are valid. Moreover, we enhance LLMs’ ability to understand the links between requirements by retrieving similar requirement pairs as in-context examples, thereby improving performance over Zero-Shot and CoT prompting. LLMs are well-suited for this task due to their ability to understand and capture subtle differences in messages and signals across variations in requirements templates, while simultaneously supporting explainability for traceability link validation.

\section{Conclusion}\label{sec:conclusion}
In this paper, we propose TVR, an approach leveraging RAG-based LLMs to verify the validity of traceability between high-level stakeholder requirements and system requirements in automotive systems. TVR achieves 98.87\% accuracy in detecting whether a traceability link is valid. Furthermore, experimental results demonstrate the robustness of TVR in effectively handling unseen variations in requirements templates, retaining 97.13\% accuracy. Additionally, TVR can identify missing links between requirements with 85.50\% correctness. These findings indicate that TVR is effective not only for traceability validation but also for recovering missing links. TVR can thus be applied to automotive systems, helping the industry to save both time and cost. In the future, the basic principles of TVR can be adapted to other types of requirements and domains where such traceability is important. 
Looking ahead, we aim to further enhance TVR's generalizability, making it applicable to a broader range of industrial scenarios.

\section{Declarations}

\subsection{Funding}
This work was supported by a research grant from Aptiv, as well as by the Discovery Grant and Canada Research Chair programs of the Natural Sciences and Engineering Research Council of Canada (NSERC), and by Research Ireland grant 13/RC/2094-2.

\subsection{Ethical Approval}
Not applicable.

\subsection{Informed Consent}
Not applicable.

\subsection{Author Contributions}
Feifei Niu led the experimental design, conducted the experiments, and drafted the manuscript. 
Rongqi Pan contributed to the experimental execution and critically reviewed the manuscript. 
Lionel C. Briand provided overall research guidance and critically reviewed the manuscript. 
Hanyang Hu contributed the dataset, provided domain-specific expertise, and critically reviewed the manuscript. 
All authors reviewed and approved the final manuscript.

\subsection{Data Availability Statement}
The replication package is publicly available on GitHub at \cite{TVR2025}.

\subsection{Conflict of Interest}
The authors declare that they have no conflict of interest.

\subsection{Clinical Trial Number}
Not applicable.

\bibliography{sn-bibliography}

\begin{thebibliography}{60}
\ifx \bisbn   \undefined \def \bisbn  #1{ISBN #1}\fi
\ifx \binits  \undefined \def \binits#1{#1}\fi
\ifx \bauthor  \undefined \def \bauthor#1{#1}\fi
\ifx \batitle  \undefined \def \batitle#1{#1}\fi
\ifx \bjtitle  \undefined \def \bjtitle#1{#1}\fi
\ifx \bvolume  \undefined \def \bvolume#1{\textbf{#1}}\fi
\ifx \byear  \undefined \def \byear#1{#1}\fi
\ifx \bissue  \undefined \def \bissue#1{#1}\fi
\ifx \bfpage  \undefined \def \bfpage#1{#1}\fi
\ifx \blpage  \undefined \def \blpage #1{#1}\fi
\ifx \burl  \undefined \def \burl#1{\textsf{#1}}\fi
\ifx \doiurl  \undefined \def \doiurl#1{\url{https://doi.org/#1}}\fi
\ifx \betal  \undefined \def \betal{\textit{et al.}}\fi
\ifx \binstitute  \undefined \def \binstitute#1{#1}\fi
\ifx \binstitutionaled  \undefined \def \binstitutionaled#1{#1}\fi
\ifx \bctitle  \undefined \def \bctitle#1{#1}\fi
\ifx \beditor  \undefined \def \beditor#1{#1}\fi
\ifx \bpublisher  \undefined \def \bpublisher#1{#1}\fi
\ifx \bbtitle  \undefined \def \bbtitle#1{#1}\fi
\ifx \bedition  \undefined \def \bedition#1{#1}\fi
\ifx \bseriesno  \undefined \def \bseriesno#1{#1}\fi
\ifx \blocation  \undefined \def \blocation#1{#1}\fi
\ifx \bsertitle  \undefined \def \bsertitle#1{#1}\fi
\ifx \bsnm \undefined \def \bsnm#1{#1}\fi
\ifx \bsuffix \undefined \def \bsuffix#1{#1}\fi
\ifx \bparticle \undefined \def \bparticle#1{#1}\fi
\ifx \barticle \undefined \def \barticle#1{#1}\fi
\bibcommenthead
\ifx \bconfdate \undefined \def \bconfdate #1{#1}\fi
\ifx \botherref \undefined \def \botherref #1{#1}\fi
\ifx \url \undefined \def \url#1{\textsf{#1}}\fi
\ifx \bchapter \undefined \def \bchapter#1{#1}\fi
\ifx \bbook \undefined \def \bbook#1{#1}\fi
\ifx \bcomment \undefined \def \bcomment#1{#1}\fi
\ifx \oauthor \undefined \def \oauthor#1{#1}\fi
\ifx \citeauthoryear \undefined \def \citeauthoryear#1{#1}\fi
\ifx \endbibitem  \undefined \def \endbibitem {}\fi
\ifx \bconflocation  \undefined \def \bconflocation#1{#1}\fi
\ifx \arxivurl  \undefined \def \arxivurl#1{\textsf{#1}}\fi
\csname PreBibitemsHook\endcsname

\bibitem[\protect\citeauthoryear{Wang et~al.}{2024}]{wang2024review}
\begin{barticle}
\bauthor{\bsnm{Wang}, \binits{W.}},
\bauthor{\bsnm{Guo}, \binits{K.}},
\bauthor{\bsnm{Cao}, \binits{W.}},
\bauthor{\bsnm{Zhu}, \binits{H.}},
\bauthor{\bsnm{Nan}, \binits{J.}},
\bauthor{\bsnm{Yu}, \binits{L.}}:
\batitle{Review of electrical and electronic architectures for autonomous vehicles: Topologies, networking and simulators}.
\bjtitle{Automotive Innovation}
\bvolume{7}(\bissue{1}),
\bfpage{82}--\blpage{101}
(\byear{2024})
\end{barticle}
\endbibitem

\bibitem[\protect\citeauthoryear{Idri and Cheikhi}{2016}]{idri2016survey}
\begin{bchapter}
\bauthor{\bsnm{Idri}, \binits{A.}},
\bauthor{\bsnm{Cheikhi}, \binits{L.}}:
\bctitle{A survey of secondary studies in software process improvement}.
In: \bbtitle{2016 IEEE/ACS 13th International Conference of Computer Systems and Applications (AICCSA)},
pp. \bfpage{1}--\blpage{8}
(\byear{2016}).
\bcomment{IEEE}
\end{bchapter}
\endbibitem

\bibitem[\protect\citeauthoryear{Garc{\'\i}a-Mireles et~al.}{2012}]{garcia2012development}
\begin{bchapter}
\bauthor{\bsnm{Garc{\'\i}a-Mireles}, \binits{G.A.}},
\bauthor{\bsnm{{\'A}ngeles~Moraga}, \binits{M.}},
\bauthor{\bsnm{Garc{\'\i}a}, \binits{F.}}:
\bctitle{Development of maturity models: a systematic literature review}.
In: \bbtitle{16th International Conference on Evaluation \& Assessment in Software Engineering (EASE 2012)},
pp. \bfpage{279}--\blpage{283}
(\byear{2012}).
\bcomment{IET}
\end{bchapter}
\endbibitem

\bibitem[\protect\citeauthoryear{{ISO/IEC 33001}}{2015-03}]{ISO33001:2015}
\begin{botherref}
\oauthor{\bsnm{{ISO/IEC 33001}}}:
{Information technology -- Process assessment -- Concepts and terminology}.
{International Organization for Standardization},
{Geneva, Switzerland}
(2015-03)
\end{botherref}
\endbibitem

\bibitem[\protect\citeauthoryear{{ISO/IEC 33002}}{2015-03}]{ISO33002:2015}
\begin{botherref}
\oauthor{\bsnm{{ISO/IEC 33002}}}:
{Information technology—Process assessment—Concepts and terminology}.
{International Organization for Standardization},
{Geneva, Switzerland}
(2015-03)
\end{botherref}
\endbibitem

\bibitem[\protect\citeauthoryear{{ISO/IEC 33003}}{2015-03}]{ISO33003:2015}
\begin{botherref}
\oauthor{\bsnm{{ISO/IEC 33003}}}:
{Information technology—Process assessment—Concepts and terminology}.
{International Organization for Standardization},
{Geneva, Switzerland}
(2015-03)
\end{botherref}
\endbibitem

\bibitem[\protect\citeauthoryear{{ISO/IEC 33004}}{2015-03}]{ISO33004:2015}
\begin{botherref}
\oauthor{\bsnm{{ISO/IEC 33004}}}:
{Information technology—Process assessment—Concepts and terminology}.
{International Organization for Standardization},
{Geneva, Switzerland}
(2015-03)
\end{botherref}
\endbibitem

\bibitem[\protect\citeauthoryear{Wiegers and Beatty}{2013}]{wiegers2013software}
\begin{bbook}
\bauthor{\bsnm{Wiegers}, \binits{K.E.}},
\bauthor{\bsnm{Beatty}, \binits{J.}}:
\bbtitle{Software Requirements}.
\bpublisher{Pearson Education}, \blocation{???}
(\byear{2013})
\end{bbook}
\endbibitem

\bibitem[\protect\citeauthoryear{Pargaonkar}{2023}]{pargaonkar2023synergizing}
\begin{barticle}
\bauthor{\bsnm{Pargaonkar}, \binits{S.}}:
\batitle{Synergizing requirements engineering and quality assurance: A comprehensive exploration in software quality engineering}.
\bjtitle{International Journal of Science and Research (IJSR)}
\bvolume{12}(\bissue{8}),
\bfpage{2003}--\blpage{2007}
(\byear{2023})
\end{barticle}
\endbibitem

\bibitem[\protect\citeauthoryear{Tufail et~al.}{2017}]{tufail2017systematic}
\begin{bchapter}
\bauthor{\bsnm{Tufail}, \binits{H.}},
\bauthor{\bsnm{Masood}, \binits{M.F.}},
\bauthor{\bsnm{Zeb}, \binits{B.}},
\bauthor{\bsnm{Azam}, \binits{F.}},
\bauthor{\bsnm{Anwar}, \binits{M.W.}}:
\bctitle{A systematic review of requirement traceability techniques and tools}.
In: \bbtitle{2017 2nd International Conference on System Reliability and Safety (ICSRS)},
pp. \bfpage{450}--\blpage{454}
(\byear{2017}).
\bcomment{IEEE}
\end{bchapter}
\endbibitem

\bibitem[\protect\citeauthoryear{Siegl et~al.}{2010}]{siegl2010model}
\begin{bchapter}
\bauthor{\bsnm{Siegl}, \binits{S.}},
\bauthor{\bsnm{Hielscher}, \binits{K.-S.}},
\bauthor{\bsnm{German}, \binits{R.}}:
\bctitle{Model based requirements analysis and testing of automotive systems with timed usage models}.
In: \bbtitle{2010 18th IEEE International Requirements Engineering Conference},
pp. \bfpage{345}--\blpage{350}
(\byear{2010}).
\bcomment{IEEE}
\end{bchapter}
\endbibitem

\bibitem[\protect\citeauthoryear{Qusef et~al.}{2011}]{qusef2011scotch}
\begin{bchapter}
\bauthor{\bsnm{Qusef}, \binits{A.}},
\bauthor{\bsnm{Bavota}, \binits{G.}},
\bauthor{\bsnm{Oliveto}, \binits{R.}},
\bauthor{\bsnm{De~Lucia}, \binits{A.}},
\bauthor{\bsnm{Binkley}, \binits{D.}}:
\bctitle{Scotch: Slicing and coupling based test to code trace hunter}.
In: \bbtitle{2011 18th Working Conference on Reverse Engineering},
pp. \bfpage{443}--\blpage{444}
(\byear{2011}).
\bcomment{IEEE}
\end{bchapter}
\endbibitem

\bibitem[\protect\citeauthoryear{Marscholik and Subke}{2009}]{marscholik2009road}
\begin{bbook}
\bauthor{\bsnm{Marscholik}, \binits{C.}},
\bauthor{\bsnm{Subke}, \binits{P.}}:
\bbtitle{Road Vehicles: Diagnostic Communication: Technology and Applications}.
\bpublisher{Laxmi Publications, Ltd.}, \blocation{???}
(\byear{2009})
\end{bbook}
\endbibitem

\bibitem[\protect\citeauthoryear{Pirasteh et~al.}{2019}]{pirasteh2019interactive}
\begin{bchapter}
\bauthor{\bsnm{Pirasteh}, \binits{P.}},
\bauthor{\bsnm{Nowaczyk}, \binits{S.}},
\bauthor{\bsnm{Pashami}, \binits{S.}},
\bauthor{\bsnm{L{\"o}wenadler}, \binits{M.}},
\bauthor{\bsnm{Thunberg}, \binits{K.}},
\bauthor{\bsnm{Ydreskog}, \binits{H.}},
\bauthor{\bsnm{Berck}, \binits{P.}}:
\bctitle{Interactive feature extraction for diagnostic trouble codes in predictive maintenance: A case study from automotive domain}.
In: \bbtitle{Proceedings of the Workshop on Interactive Data Mining},
pp. \bfpage{1}--\blpage{10}
(\byear{2019})
\end{bchapter}
\endbibitem

\bibitem[\protect\citeauthoryear{Gao et~al.}{2022}]{gao2022using}
\begin{bchapter}
\bauthor{\bsnm{Gao}, \binits{H.}},
\bauthor{\bsnm{Kuang}, \binits{H.}},
\bauthor{\bsnm{Sun}, \binits{K.}},
\bauthor{\bsnm{Ma}, \binits{X.}},
\bauthor{\bsnm{Egyed}, \binits{A.}},
\bauthor{\bsnm{M{\"a}der}, \binits{P.}},
\bauthor{\bsnm{Rong}, \binits{G.}},
\bauthor{\bsnm{Shao}, \binits{D.}},
\bauthor{\bsnm{Zhang}, \binits{H.}}:
\bctitle{Using consensual biterms from text structures of requirements and code to improve ir-based traceability recovery}.
In: \bbtitle{Proceedings of the 37th IEEE/ACM International Conference on Automated Software Engineering},
pp. \bfpage{1}--\blpage{1}
(\byear{2022})
\end{bchapter}
\endbibitem

\bibitem[\protect\citeauthoryear{Hey et~al.}{2021}]{hey2021improving}
\begin{bchapter}
\bauthor{\bsnm{Hey}, \binits{T.}},
\bauthor{\bsnm{Chen}, \binits{F.}},
\bauthor{\bsnm{Weigelt}, \binits{S.}},
\bauthor{\bsnm{Tichy}, \binits{W.F.}}:
\bctitle{Improving traceability link recovery using fine-grained requirements-to-code relations}.
In: \bbtitle{2021 IEEE International Conference on Software Maintenance and Evolution (ICSME)},
pp. \bfpage{12}--\blpage{22}
(\byear{2021}).
\bcomment{IEEE}
\end{bchapter}
\endbibitem

\bibitem[\protect\citeauthoryear{Panichella et~al.}{2013}]{panichella2013and}
\begin{bchapter}
\bauthor{\bsnm{Panichella}, \binits{A.}},
\bauthor{\bsnm{McMillan}, \binits{C.}},
\bauthor{\bsnm{Moritz}, \binits{E.}},
\bauthor{\bsnm{Palmieri}, \binits{D.}},
\bauthor{\bsnm{Oliveto}, \binits{R.}},
\bauthor{\bsnm{Poshyvanyk}, \binits{D.}},
\bauthor{\bsnm{De~Lucia}, \binits{A.}}:
\bctitle{When and how using structural information to improve ir-based traceability recovery}.
In: \bbtitle{2013 17th European Conference on Software Maintenance and Reengineering},
pp. \bfpage{199}--\blpage{208}
(\byear{2013}).
\bcomment{IEEE}
\end{bchapter}
\endbibitem

\bibitem[\protect\citeauthoryear{Kuang et~al.}{2015}]{kuang2015can}
\begin{barticle}
\bauthor{\bsnm{Kuang}, \binits{H.}},
\bauthor{\bsnm{M{\"a}der}, \binits{P.}},
\bauthor{\bsnm{Hu}, \binits{H.}},
\bauthor{\bsnm{Ghabi}, \binits{A.}},
\bauthor{\bsnm{Huang}, \binits{L.}},
\bauthor{\bsnm{L{\"u}}, \binits{J.}},
\bauthor{\bsnm{Egyed}, \binits{A.}}:
\batitle{Can method data dependencies support the assessment of traceability between requirements and source code?}
\bjtitle{Journal of Software: Evolution and Process}
\bvolume{27}(\bissue{11}),
\bfpage{838}--\blpage{866}
(\byear{2015})
\end{barticle}
\endbibitem

\bibitem[\protect\citeauthoryear{Antoniol et~al.}{2002}]{antoniol2002recovering}
\begin{barticle}
\bauthor{\bsnm{Antoniol}, \binits{G.}},
\bauthor{\bsnm{Canfora}, \binits{G.}},
\bauthor{\bsnm{Casazza}, \binits{G.}},
\bauthor{\bsnm{De~Lucia}, \binits{A.}},
\bauthor{\bsnm{Merlo}, \binits{E.}}:
\batitle{Recovering traceability links between code and documentation}.
\bjtitle{IEEE transactions on software engineering}
\bvolume{28}(\bissue{10}),
\bfpage{970}--\blpage{983}
(\byear{2002})
\end{barticle}
\endbibitem

\bibitem[\protect\citeauthoryear{Marcus and Maletic}{2003}]{marcus2003recovering}
\begin{bchapter}
\bauthor{\bsnm{Marcus}, \binits{A.}},
\bauthor{\bsnm{Maletic}, \binits{J.I.}}:
\bctitle{Recovering documentation-to-source-code traceability links using latent semantic indexing}.
In: \bbtitle{25th International Conference on Software Engineering, 2003. Proceedings.},
pp. \bfpage{125}--\blpage{135}
(\byear{2003}).
\bcomment{IEEE}
\end{bchapter}
\endbibitem

\bibitem[\protect\citeauthoryear{Keim et~al.}{2024}]{keim2024recovering}
\begin{bchapter}
\bauthor{\bsnm{Keim}, \binits{J.}},
\bauthor{\bsnm{Corallo}, \binits{S.}},
\bauthor{\bsnm{Fuch{\ss}}, \binits{D.}},
\bauthor{\bsnm{Hey}, \binits{T.}},
\bauthor{\bsnm{Telge}, \binits{T.}},
\bauthor{\bsnm{Koziolek}, \binits{A.}}:
\bctitle{Recovering trace links between software documentation and code}.
In: \bbtitle{Proceedings of the IEEE/ACM 46th International Conference on Software Engineering},
pp. \bfpage{1}--\blpage{13}
(\byear{2024})
\end{bchapter}
\endbibitem

\bibitem[\protect\citeauthoryear{Keim et~al.}{2023}]{keim2023detecting}
\begin{bchapter}
\bauthor{\bsnm{Keim}, \binits{J.}},
\bauthor{\bsnm{Corallo}, \binits{S.}},
\bauthor{\bsnm{Fuch{\ss}}, \binits{D.}},
\bauthor{\bsnm{Koziolek}, \binits{A.}}:
\bctitle{Detecting inconsistencies in software architecture documentation using traceability link recovery}.
In: \bbtitle{2023 IEEE 20th International Conference on Software Architecture (ICSA)},
pp. \bfpage{141}--\blpage{152}
(\byear{2023}).
\bcomment{IEEE}
\end{bchapter}
\endbibitem

\bibitem[\protect\citeauthoryear{Cleland-Huang et~al.}{2005}]{cleland2005utilizing}
\begin{bchapter}
\bauthor{\bsnm{Cleland-Huang}, \binits{J.}},
\bauthor{\bsnm{Settimi}, \binits{R.}},
\bauthor{\bsnm{Duan}, \binits{C.}},
\bauthor{\bsnm{Zou}, \binits{X.}}:
\bctitle{Utilizing supporting evidence to improve dynamic requirements traceability}.
In: \bbtitle{13th IEEE International Conference on Requirements Engineering (RE'05)},
pp. \bfpage{135}--\blpage{144}
(\byear{2005}).
\bcomment{IEEE}
\end{bchapter}
\endbibitem

\bibitem[\protect\citeauthoryear{Brown et~al.}{2020}]{brown2020language}
\begin{barticle}
\bauthor{\bsnm{Brown}, \binits{T.}},
\bauthor{\bsnm{Mann}, \binits{B.}},
\bauthor{\bsnm{Ryder}, \binits{N.}},
\bauthor{\bsnm{Subbiah}, \binits{M.}},
\bauthor{\bsnm{Kaplan}, \binits{J.D.}},
\bauthor{\bsnm{Dhariwal}, \binits{P.}},
\bauthor{\bsnm{Neelakantan}, \binits{A.}},
\bauthor{\bsnm{Shyam}, \binits{P.}},
\bauthor{\bsnm{Sastry}, \binits{G.}},
\bauthor{\bsnm{Askell}, \binits{A.}}, \betal:
\batitle{Language models are few-shot learners}.
\bjtitle{Advances in neural information processing systems}
\bvolume{33},
\bfpage{1877}--\blpage{1901}
(\byear{2020})
\end{barticle}
\endbibitem

\bibitem[\protect\citeauthoryear{Qin et~al.}{2023}]{qin2023chatgpt}
\begin{botherref}
\oauthor{\bsnm{Qin}, \binits{C.}},
\oauthor{\bsnm{Zhang}, \binits{A.}},
\oauthor{\bsnm{Zhang}, \binits{Z.}},
\oauthor{\bsnm{Chen}, \binits{J.}},
\oauthor{\bsnm{Yasunaga}, \binits{M.}},
\oauthor{\bsnm{Yang}, \binits{D.}}:
Is chatgpt a general-purpose natural language processing task solver?
arXiv preprint arXiv:2302.06476
(2023)
\end{botherref}
\endbibitem

\bibitem[\protect\citeauthoryear{Fuch{\ss} et~al.}{2025}]{fuchss2025lissa}
\begin{bchapter}
\bauthor{\bsnm{Fuch{\ss}}, \binits{D.}},
\bauthor{\bsnm{Hey}, \binits{T.}},
\bauthor{\bsnm{Keim}, \binits{J.}},
\bauthor{\bsnm{Liu}, \binits{H.}},
\bauthor{\bsnm{Ewald}, \binits{N.}},
\bauthor{\bsnm{Thirolf}, \binits{T.}},
\bauthor{\bsnm{Koziolek}, \binits{A.}}:
\bctitle{Lissa: Toward generic traceability link recovery through retrieval-augmented generation}.
In: \bbtitle{Proceedings of the IEEE/ACM 47th International Conference on Software Engineering. ICSE},
vol. \bseriesno{25}
(\byear{2025})
\end{bchapter}
\endbibitem

\bibitem[\protect\citeauthoryear{Rodriguez et~al.}{2023}]{rodriguez2023prompts}
\begin{bchapter}
\bauthor{\bsnm{Rodriguez}, \binits{A.D.}},
\bauthor{\bsnm{Dearstyne}, \binits{K.R.}},
\bauthor{\bsnm{Cleland-Huang}, \binits{J.}}:
\bctitle{Prompts matter: Insights and strategies for prompt engineering in automated software traceability}.
In: \bbtitle{2023 IEEE 31st International Requirements Engineering Conference Workshops (REW)},
pp. \bfpage{455}--\blpage{464}
(\byear{2023}).
\bcomment{IEEE}
\end{bchapter}
\endbibitem

\bibitem[\protect\citeauthoryear{Lewis et~al.}{2020}]{lewis2020retrieval}
\begin{barticle}
\bauthor{\bsnm{Lewis}, \binits{P.}},
\bauthor{\bsnm{Perez}, \binits{E.}},
\bauthor{\bsnm{Piktus}, \binits{A.}},
\bauthor{\bsnm{Petroni}, \binits{F.}},
\bauthor{\bsnm{Karpukhin}, \binits{V.}},
\bauthor{\bsnm{Goyal}, \binits{N.}},
\bauthor{\bsnm{K{\"u}ttler}, \binits{H.}},
\bauthor{\bsnm{Lewis}, \binits{M.}},
\bauthor{\bsnm{Yih}, \binits{W.-t.}},
\bauthor{\bsnm{Rockt{\"a}schel}, \binits{T.}}, \betal:
\batitle{Retrieval-augmented generation for knowledge-intensive nlp tasks}.
\bjtitle{Advances in neural information processing systems}
\bvolume{33},
\bfpage{9459}--\blpage{9474}
(\byear{2020})
\end{barticle}
\endbibitem

\bibitem[\protect\citeauthoryear{Feifei et~al.}{2025}]{TVR2025}
\begin{botherref}
\oauthor{\bsnm{Feifei}, \binits{N.}},
\oauthor{\bsnm{Rongqi}, \binits{P.}},
\oauthor{\bsnm{Lionel~C.}, \binits{B.}},
\oauthor{\bsnm{Hanyang}, \binits{H.}}:
TVR.
\url{https://github.com/feifeiniu-se/TVR}.
Accessed: 2025-09-28
(2025)
\end{botherref}
\endbibitem

\bibitem[\protect\citeauthoryear{Theissler}{2017}]{theissler2017multi}
\begin{bchapter}
\bauthor{\bsnm{Theissler}, \binits{A.}}:
\bctitle{Multi-class novelty detection in diagnostic trouble codes from repair shops}.
In: \bbtitle{2017 IEEE 15th International Conference on Industrial Informatics (INDIN)},
pp. \bfpage{1043}--\blpage{1049}
(\byear{2017}).
\bcomment{IEEE}
\end{bchapter}
\endbibitem

\bibitem[\protect\citeauthoryear{Palai}{2013}]{palai2013vehicle}
\begin{botherref}
\oauthor{\bsnm{Palai}, \binits{D.}}:
Vehicle level approach for optimization of on-board diagnostic strategies for fault management
(2013)
\end{botherref}
\endbibitem

\bibitem[\protect\citeauthoryear{Gotel and Finkelstein}{1994}]{gotel1994analysis}
\begin{bchapter}
\bauthor{\bsnm{Gotel}, \binits{O.C.}},
\bauthor{\bsnm{Finkelstein}, \binits{C.}}:
\bctitle{An analysis of the requirements traceability problem}.
In: \bbtitle{Proceedings of IEEE International Conference on Requirements Engineering},
pp. \bfpage{94}--\blpage{101}
(\byear{1994}).
\bcomment{IEEE}
\end{bchapter}
\endbibitem

\bibitem[\protect\citeauthoryear{Rahimi and Cleland-Huang}{2018}]{rahimi2018evolving}
\begin{barticle}
\bauthor{\bsnm{Rahimi}, \binits{M.}},
\bauthor{\bsnm{Cleland-Huang}, \binits{J.}}:
\batitle{Evolving software trace links between requirements and source code}.
\bjtitle{Empirical Software Engineering}
\bvolume{23},
\bfpage{2198}--\blpage{2231}
(\byear{2018})
\end{barticle}
\endbibitem

\bibitem[\protect\citeauthoryear{Charalampidou et~al.}{2021}]{charalampidou2021empirical}
\begin{barticle}
\bauthor{\bsnm{Charalampidou}, \binits{S.}},
\bauthor{\bsnm{Ampatzoglou}, \binits{A.}},
\bauthor{\bsnm{Karountzos}, \binits{E.}},
\bauthor{\bsnm{Avgeriou}, \binits{P.}}:
\batitle{Empirical studies on software traceability: A mapping study}.
\bjtitle{Journal of Software: Evolution and Process}
\bvolume{33}(\bissue{2}),
\bfpage{2294}
(\byear{2021})
\end{barticle}
\endbibitem

\bibitem[\protect\citeauthoryear{Guo et~al.}{2024}]{guo2024natural}
\begin{botherref}
\oauthor{\bsnm{Guo}, \binits{J.L.}},
\oauthor{\bsnm{Stegh{\"o}fer}, \binits{J.-P.}},
\oauthor{\bsnm{Vogelsang}, \binits{A.}},
\oauthor{\bsnm{Cleland-Huang}, \binits{J.}}:
Natural language processing for requirements traceability.
arXiv preprint arXiv:2405.10845
(2024)
\end{botherref}
\endbibitem

\bibitem[\protect\citeauthoryear{{Amazon Web Services}}{2025}]{aws_titan}
\begin{botherref}
\oauthor{\bsnm{{Amazon Web Services}}}:
Amazon Titan Embedding Models -- AWS Bedrock.
\url{https://docs.aws.amazon.com/bedrock/latest/userguide/titan-embedding-models.html}.
Accessed: 2025-09-29
(2025)
\end{botherref}
\endbibitem

\bibitem[\protect\citeauthoryear{{Facebook Research}}{2025}]{faiss}
\begin{botherref}
\oauthor{\bsnm{{Facebook Research}}}:
Faiss: A library for efficient similarity search and clustering of dense vectors.
\url{https://github.com/facebookresearch/faiss}.
Accessed: 2025-09-29
(2025)
\end{botherref}
\endbibitem

\bibitem[\protect\citeauthoryear{{Amazon Web Services}}{2025}]{aws_anthropic}
\begin{botherref}
\oauthor{\bsnm{{Amazon Web Services}}}:
Anthropic Claude on AWS Bedrock.
\url{https://aws.amazon.com/bedrock/anthropic/}.
Accessed: 2025-09-29
(2025)
\end{botherref}
\endbibitem

\bibitem[\protect\citeauthoryear{{Anthropic}}{2025}]{claude_prompt}
\begin{botherref}
\oauthor{\bsnm{{Anthropic}}}:
Prompt Engineering with Claude: Use XML Tags.
\url{https://docs.claude.com/en/docs/build-with-claude/prompt-engineering/use-xml-tags}.
Accessed: 2025-09-29
(2025)
\end{botherref}
\endbibitem

\bibitem[\protect\citeauthoryear{Hastie et~al.}{2009}]{hastie2009elements}
\begin{bbook}
\bauthor{\bsnm{Hastie}, \binits{T.}},
\bauthor{\bsnm{Tibshirani}, \binits{R.}},
\bauthor{\bsnm{Friedman}, \binits{J.H.}},
\bauthor{\bsnm{Friedman}, \binits{J.H.}}:
\bbtitle{The Elements of Statistical Learning: Data Mining, Inference, and Prediction}
vol. \bseriesno{2}.
\bpublisher{Springer}, \blocation{???}
(\byear{2009})
\end{bbook}
\endbibitem

\bibitem[\protect\citeauthoryear{Sch{\"u}tze et~al.}{2008}]{schutze2008introduction}
\begin{bbook}
\bauthor{\bsnm{Sch{\"u}tze}, \binits{H.}},
\bauthor{\bsnm{Manning}, \binits{C.D.}},
\bauthor{\bsnm{Raghavan}, \binits{P.}}:
\bbtitle{Introduction to Information Retrieval}
vol. \bseriesno{39}.
\bpublisher{Cambridge University Press Cambridge}, \blocation{???}
(\byear{2008})
\end{bbook}
\endbibitem

\bibitem[\protect\citeauthoryear{Reimers and Gurevych}{2019}]{reimers2019sentence}
\begin{botherref}
\oauthor{\bsnm{Reimers}, \binits{N.}},
\oauthor{\bsnm{Gurevych}, \binits{I.}}:
Sentence-bert: Sentence embeddings using siamese bert-networks.
arXiv preprint arXiv:1908.10084
(2019)
\end{botherref}
\endbibitem

\bibitem[\protect\citeauthoryear{Wang et~al.}{2022}]{wang2022self}
\begin{botherref}
\oauthor{\bsnm{Wang}, \binits{X.}},
\oauthor{\bsnm{Wei}, \binits{J.}},
\oauthor{\bsnm{Schuurmans}, \binits{D.}},
\oauthor{\bsnm{Le}, \binits{Q.}},
\oauthor{\bsnm{Chi}, \binits{E.}},
\oauthor{\bsnm{Narang}, \binits{S.}},
\oauthor{\bsnm{Chowdhery}, \binits{A.}},
\oauthor{\bsnm{Zhou}, \binits{D.}}:
Self-consistency improves chain of thought reasoning in language models.
arXiv preprint arXiv:2203.11171
(2022)
\end{botherref}
\endbibitem

\bibitem[\protect\citeauthoryear{Lucia et~al.}{2007}]{lucia2007recovering}
\begin{barticle}
\bauthor{\bsnm{Lucia}, \binits{A.D.}},
\bauthor{\bsnm{Fasano}, \binits{F.}},
\bauthor{\bsnm{Oliveto}, \binits{R.}},
\bauthor{\bsnm{Tortora}, \binits{G.}}:
\batitle{Recovering traceability links in software artifact management systems using information retrieval methods}.
\bjtitle{ACM Transactions on Software Engineering and Methodology (TOSEM)}
\bvolume{16}(\bissue{4}),
\bfpage{13}
(\byear{2007})
\end{barticle}
\endbibitem

\bibitem[\protect\citeauthoryear{Mahmoud}{2015}]{mahmoud2015information}
\begin{bchapter}
\bauthor{\bsnm{Mahmoud}, \binits{A.}}:
\bctitle{An information theoretic approach for extracting and tracing non-functional requirements}.
In: \bbtitle{2015 IEEE 23rd International Requirements Engineering Conference (RE)},
pp. \bfpage{36}--\blpage{45}
(\byear{2015}).
\bcomment{IEEE}
\end{bchapter}
\endbibitem

\bibitem[\protect\citeauthoryear{Asuncion et~al.}{2010}]{asuncion2010software}
\begin{bchapter}
\bauthor{\bsnm{Asuncion}, \binits{H.U.}},
\bauthor{\bsnm{Asuncion}, \binits{A.U.}},
\bauthor{\bsnm{Taylor}, \binits{R.N.}}:
\bctitle{Software traceability with topic modeling}.
In: \bbtitle{Proceedings of the 32nd ACM/IEEE International Conference on Software Engineering-Volume 1},
pp. \bfpage{95}--\blpage{104}
(\byear{2010})
\end{bchapter}
\endbibitem

\bibitem[\protect\citeauthoryear{Gethers et~al.}{2011}]{gethers2011integrating}
\begin{bchapter}
\bauthor{\bsnm{Gethers}, \binits{M.}},
\bauthor{\bsnm{Oliveto}, \binits{R.}},
\bauthor{\bsnm{Poshyvanyk}, \binits{D.}},
\bauthor{\bsnm{De~Lucia}, \binits{A.}}:
\bctitle{On integrating orthogonal information retrieval methods to improve traceability recovery}.
In: \bbtitle{2011 27th IEEE International Conference on Software Maintenance (ICSM)},
pp. \bfpage{133}--\blpage{142}
(\byear{2011}).
\bcomment{IEEE}
\end{bchapter}
\endbibitem

\bibitem[\protect\citeauthoryear{Moran et~al.}{2020}]{moran2020improving}
\begin{bchapter}
\bauthor{\bsnm{Moran}, \binits{K.}},
\bauthor{\bsnm{Palacio}, \binits{D.N.}},
\bauthor{\bsnm{Bernal-C{\'a}rdenas}, \binits{C.}},
\bauthor{\bsnm{McCrystal}, \binits{D.}},
\bauthor{\bsnm{Poshyvanyk}, \binits{D.}},
\bauthor{\bsnm{Shenefiel}, \binits{C.}},
\bauthor{\bsnm{Johnson}, \binits{J.}}:
\bctitle{Improving the effectiveness of traceability link recovery using hierarchical bayesian networks}.
In: \bbtitle{Proceedings of the ACM/IEEE 42nd International Conference on Software Engineering},
pp. \bfpage{873}--\blpage{885}
(\byear{2020})
\end{bchapter}
\endbibitem

\bibitem[\protect\citeauthoryear{Guo et~al.}{2017}]{guo2017semantically}
\begin{bchapter}
\bauthor{\bsnm{Guo}, \binits{J.}},
\bauthor{\bsnm{Cheng}, \binits{J.}},
\bauthor{\bsnm{Cleland-Huang}, \binits{J.}}:
\bctitle{Semantically enhanced software traceability using deep learning techniques}.
In: \bbtitle{2017 IEEE/ACM 39th International Conference on Software Engineering (ICSE)},
pp. \bfpage{3}--\blpage{14}
(\byear{2017}).
\bcomment{IEEE}
\end{bchapter}
\endbibitem

\bibitem[\protect\citeauthoryear{Wang et~al.}{2018}]{wang2018enhancing}
\begin{bchapter}
\bauthor{\bsnm{Wang}, \binits{W.}},
\bauthor{\bsnm{Niu}, \binits{N.}},
\bauthor{\bsnm{Liu}, \binits{H.}},
\bauthor{\bsnm{Niu}, \binits{Z.}}:
\bctitle{Enhancing automated requirements traceability by resolving polysemy}.
In: \bbtitle{2018 IEEE 26th International Requirements Engineering Conference (RE)},
pp. \bfpage{40}--\blpage{51}
(\byear{2018}).
\bcomment{IEEE}
\end{bchapter}
\endbibitem

\bibitem[\protect\citeauthoryear{Zhao et~al.}{2017}]{zhao2017improved}
\begin{bchapter}
\bauthor{\bsnm{Zhao}, \binits{T.}},
\bauthor{\bsnm{Cao}, \binits{Q.}},
\bauthor{\bsnm{Sun}, \binits{Q.}}:
\bctitle{An improved approach to traceability recovery based on word embeddings}.
In: \bbtitle{2017 24th Asia-Pacific Software Engineering Conference (APSEC)},
pp. \bfpage{81}--\blpage{89}
(\byear{2017}).
\bcomment{IEEE}
\end{bchapter}
\endbibitem

\bibitem[\protect\citeauthoryear{Mills et~al.}{2019}]{mills2019tracing}
\begin{bchapter}
\bauthor{\bsnm{Mills}, \binits{C.}},
\bauthor{\bsnm{Escobar-Avila}, \binits{J.}},
\bauthor{\bsnm{Bhattacharya}, \binits{A.}},
\bauthor{\bsnm{Kondyukov}, \binits{G.}},
\bauthor{\bsnm{Chakraborty}, \binits{S.}},
\bauthor{\bsnm{Haiduc}, \binits{S.}}:
\bctitle{Tracing with less data: active learning for classification-based traceability link recovery}.
In: \bbtitle{2019 IEEE International Conference on Software Maintenance and Evolution (ICSME)},
pp. \bfpage{103}--\blpage{113}
(\byear{2019}).
\bcomment{IEEE}
\end{bchapter}
\endbibitem

\bibitem[\protect\citeauthoryear{Zhang et~al.}{2021}]{zhang2021recovering}
\begin{bchapter}
\bauthor{\bsnm{Zhang}, \binits{M.}},
\bauthor{\bsnm{Tao}, \binits{C.}},
\bauthor{\bsnm{Guo}, \binits{H.}},
\bauthor{\bsnm{Huang}, \binits{Z.}}:
\bctitle{Recovering semantic traceability between requirements and source code using feature representation techniques}.
In: \bbtitle{2021 IEEE 21st International Conference on Software Quality, Reliability and Security (QRS)},
pp. \bfpage{873}--\blpage{882}
(\byear{2021}).
\bcomment{IEEE}
\end{bchapter}
\endbibitem

\bibitem[\protect\citeauthoryear{OpenAI}{2023a}]{openai2023chatgpt}
\begin{botherref}
\oauthor{\bsnm{OpenAI}}:
ChatGPT
(2023).
\url{https://openai.com/chatgpt}
\end{botherref}
\endbibitem

\bibitem[\protect\citeauthoryear{OpenAI}{2023b}]{openai2023gpt4}
\begin{botherref}
\oauthor{\bsnm{OpenAI}}:
GPT-4 Technical Report
(2023).
\url{https://arxiv.org/abs/2303.08774}
\end{botherref}
\endbibitem

\bibitem[\protect\citeauthoryear{https://www.llama.com/}{2023}]{llama}
\begin{botherref}
\oauthor{\bsnm{https://www.llama.com/}}:
Llama
(2023).
\url{https://www.llama.com/}
\end{botherref}
\endbibitem

\bibitem[\protect\citeauthoryear{https://claude.ai/}{2023}]{claude}
\begin{botherref}
\oauthor{\bsnm{https://claude.ai/}}:
Claude
(2023).
\url{https://claude.ai/}
\end{botherref}
\endbibitem

\bibitem[\protect\citeauthoryear{Kusner et~al.}{2015}]{kusner2015word}
\begin{bchapter}
\bauthor{\bsnm{Kusner}, \binits{M.}},
\bauthor{\bsnm{Sun}, \binits{Y.}},
\bauthor{\bsnm{Kolkin}, \binits{N.}},
\bauthor{\bsnm{Weinberger}, \binits{K.}}:
\bctitle{From word embeddings to document distances}.
In: \bbtitle{International Conference on Machine Learning},
pp. \bfpage{957}--\blpage{966}
(\byear{2015}).
\bcomment{PMLR}
\end{bchapter}
\endbibitem

\bibitem[\protect\citeauthoryear{Hey et~al.}{2024}]{hey2024requirements}
\begin{bchapter}
\bauthor{\bsnm{Hey}, \binits{T.}},
\bauthor{\bsnm{Keim}, \binits{J.}},
\bauthor{\bsnm{Corallo}, \binits{S.}}:
\bctitle{Requirements classification for traceability link recovery}.
In: \bbtitle{2024 IEEE 32nd International Requirements Engineering Conference (RE)},
pp. \bfpage{155}--\blpage{167}
(\byear{2024}).
\bcomment{IEEE}
\end{bchapter}
\endbibitem

\bibitem[\protect\citeauthoryear{Hey et~al.}{2025}]{hey2025requirements}
\begin{bchapter}
\bauthor{\bsnm{Hey}, \binits{T.}},
\bauthor{\bsnm{Fuch{\ss}}, \binits{D.}},
\bauthor{\bsnm{Keim}, \binits{J.}},
\bauthor{\bsnm{Koziolek}, \binits{A.}}:
\bctitle{Requirements traceability link recovery via retrieval-augmented generation}.
In: \bbtitle{International Working Conference on Requirements Engineering: Foundation for Software Quality},
pp. \bfpage{381}--\blpage{397}
(\byear{2025}).
\bcomment{Springer}
\end{bchapter}
\endbibitem

\end{thebibliography}

\end{document}